\documentclass[useAMS,usenatbib]{mn2e}
\DeclareSymbolFont{cmletters}{OML}{cmm}{m}{it}
\DeclareMathSymbol{v}{\mathalpha}{cmletters}{"76}

\usepackage{amsmath}
\usepackage{amssymb}
\usepackage{epsfig}
\usepackage{graphicx}
\usepackage{ifthen}
\usepackage{latexsym}
\usepackage{rotating}
\usepackage{subfigure}
\usepackage{times,epsf}
\usepackage{txfonts}
\usepackage{varioref}
\usepackage{verbatim}
\usepackage{array}
\usepackage{url}
\usepackage{color}
\usepackage[dvipsnames]{xcolor}
\usepackage[T1]{fontenc}

\newcommand{\be}{\begin{equation}}
\newcommand{\ee}{\end{equation}}
\newcommand{\bea}{\begin{eqnarray}}
\newcommand{\eea}{\end{eqnarray}}
\newcommand{\gdet}{\sqrt{-g}}

\newcommand\apj{Astrophysical Journal}
\newcommand\apjl{Astrophysical Journal Letters}

\newcommand\apjs{Astrophysical Journal Suppl. Ser.}

\newcommand\aap{Astronomy \& Astrophysics}

\newcommand\nat{Nature}

\newcommand\mnras{Monthly Notices of the Royal Astronomical Society}

\newcommand\pasj{Publications of the Astronomical Society of Japan}

\newcommand\jqsrt{Journal of Quantitative Spectroscopy and Radiative Transfer}

\newcommand{\avg}[1]{\langle #1\rangle}

\newcommand{\koral}{\texttt{KORAL}}
\newcommand{\Medd}{\dot M_{\rm Edd}}

\title[Super-critical accretion disks]{
Numerical simulations of super-critical black hole accretion flows in general relativity}
\author[A. S\k{a}dowski, R. Narayan, J. C. McKinney, A. Tchekhovskoy]
       {Aleksander S\k{a}dowski$^1$\footnotemark[1], 
	Ramesh Narayan$^{1}$\footnotemark[1], 
        Jonathan C. McKinney$^{2}$\footnotemark[1]
        \newauthor
        and Alexander Tchekhovskoy$^{3}$\thanks{E-mail: asadowski@cfa.harvard.edu (AS); 
rnarayan@cfa.harvard.edu (RN);	atchekho@princeton.edu (AT);  yzhu@cfa.harvard.edu (YZ);} \\
        $^1$ Harvard-Smithsonian Center for Astrophysics, 60 Garden St., Cambridge, MA 02134, USA\\
        $^2$ University of Maryland at College Park, Dept. of Physics, Joint Space-Science Institute, 1117 John S. Toll Building \#082, College Park, MD 20742, USA \\
        $^3$ Lawrence Berkeley National Laboratory, 1 Cyclotron Rd, Berkeley, CA 94720, USA; Einstein Fellow}

\begin{document}

\maketitle

\label{firstpage}

\begin{abstract}

A new general relativistic radiation magnetohydrodynamical code \koral\,
is described, which employs the M1 scheme to close the radiation
moment equations. The code has been successfully verified against a
number of tests. Axisymmetric simulations of super-critical magnetized
accretion on a non-rotating black hole ($a_*=0.0$) and a spinning
black hole ($a_*=0.9$) are presented. The accretion rates in the two
models are $\dot M\approx 100\div200 \dot M_{\rm Edd}$. These first
general relativistic simulations of super-critical black hole
accretion are potentially relevant to tidal disruption events and
hyper-accreting supermassive black holes in the early universe. Both
simulated models are optically and geometrically thick, and have
funnels through which energy escapes in the form of relativistic gas,
Poynting flux and radiative flux. The jet is significantly more
powerful in the $a_*=0.9$ run. The net energy outflow rate in the two
runs correspond to efficiencies of 5\% ($a_*=0$) and 33\% ($a_*=0.9$),
as measured with respect to the mass accretion rate at the black
hole. These efficiencies agree well with those measured in previous
simulations of non-radiative geometrically thick disks. Furthermore,
in the $a_*=0.9$ run, the outflow power appears to originate in the
spinning black hole, suggesting that the associated physics is again
similar in non-radiative and super-critical accretion flows. While the
two simulations are efficient in terms of total energy outflow, both
runs are radiatively inefficient. Their luminosities are only
$\sim1-10 L_{\rm Edd}$, which corresponds to a radiative efficiency
$\sim 0.1\%$. Interestingly, most of the radiative luminosity
emerges through the funnels, which subtend a very small solid
angle. Therefore, measured in terms of a local radiative flux, the
emitted radiation is highly super-Eddington.

\end{abstract}


\section{Introduction}
\label{introduction}

Accretion onto black holes (BHs) can occur through two major modes. If the amount of
mass reaching the BH in a given time (mass accretion rate, $\dot M$) is small, the
accretion disk is expected to be optically thin. Such a disk cannot cool
efficiently. Therefore, it is hot and geometrically thick \citep{narayanmcclintock-08,yuannarayan+14}. This mode
of accretion is observed in quiescent AGN and in the low/hard state of
X-ray binaries. Because of the relative simplicity of the physics
involved (cooling is local and in many cases negligible), such systems
have been extensively studied numerically. Having a code evolving a
relativistic gas threaded by seed magnetic field in the proper general
relativistic metric
is enough to obtain the disk structure, at least in the very vicinity
of the BH. In the last decade a number of such codes were developed
\citep[e.g.,][]{devilliersetal03,gammie03,anninosetal05,delzannaetal07}.
They were used to study, e.g., physics of accretion near the inner edge
of the disk \citep{shafee08,penna10,nobleetal11}, jets
\citep[e.g.,][]{tchekh10b,tchekh10a,tchekh11,tchekh+12,narayan+10,penna+13b}, 
outflows \citep[e.g.,][]{narayan+12b,sadowski+outflows},
viscosity \citep[e.g.,][]{penna+13a}. What is more, the small
optical depth allows us to
calculate the electromagnetic spectrum in a separate postprocessing step
\citep[e.g.,][]{shcherbakovetal10,dexterfragile-11}.  All these
factors make hot thick disks relatively well understood.

The other accretion mode is optically thick accretion. In principle,
it could take place at any accretion rate \citep{ss73}. For the
smallest rates the disk cools efficiently and is geometrically
thin. It can be either gas or radiation pressure dominated. When the
accretion rate approaches and exceeds the Eddington rate,
\be
\label{e.mdotedd}
\dot M_{\rm  Edd}=\frac1\eta\frac1{c^2}L_{\rm Edd}=\frac1\eta\frac1{c^2}\frac{4\pi GMc}{\kappa_{\rm es}}=
2.44\times 10^{18} \frac{M}{M_{\odot}} \,\rm g/s
\ee 
(where we put the efficiency of a radiativelly efficient
thin disk around a non-rotating BH\footnote{Such
definition of $\dot M_{\rm Edd}$ ensures that a thin disk with $a_*=0.0$ and
$\dot M=\dot M_{\rm Edd}$ emits the Eddington luminosity, $L_{\rm Edd}$. However, this correspondence
between the Eddington accretion rate and the Eddington luminosity 
is no longer valid for $a_*>0$ because efficiency of accretion on
rotating BHs is higher. For example, a thin disk around a BH with $a_*=0.9$ ($\eta=0.11$) 
emits $1.93 L_{\rm Edd}$.}, $\eta=0.057$,
and $\kappa_{\rm es}=0.4\, \rm cm^2/g$), photons do not have enough time to diffuse out
 of the disk. Instead, they are advected with the flow, and end up
inside the BH \citep{begelman78,abra88}.  Such photon-trapped accretion is,
as in the case of hot thick disks, radiatively inefficient.

Optically thick accretion is understood only superficially. Most of our
knowledge and modeling of such disks is based on the 40 year old
papers of \cite{ss73} and \cite{novikovthorne73}, who constructed a
one-dimensional (1D) model of a radiatively
efficient disk. Since that time, a large amount of data has been obtained by
optical, X-ray, and $\gamma$-ray observatories.  This new data has only
brought new questions which remain open even today.

In contrast to optically thin disks, numerical studies of optically
thick disks have so far been almost \textit{terra incognita}. This is
because of the level of complexity that the radiation field introduces ---
when scattering 
opacity is not negligible, each photon is scattered many times between 
its origin and the point of absorption, and 
tracking of single photons,
is too demanding even for grey (frequency independent) problems. Other 
approaches must be considered. Most useful is to treat the radiation field
as another fluid described by its own stress-energy tensor. The exact form of this tensor
depends on the adopted approximation --- the closure scheme. 

The only existing radiative global simulations of accretion disks have
been performed with non-relativistic codes and using the flux-limited
diffusion or the Eddington approximation
\citep{ohsuga09,ohsuga11}. These approaches are far from satisfactory
as they do not handle optically thin regions, e.g., disk corona,
properly, and do not include general relativity (GR). Recently, a more sophisticated approach --- the variable
Eddington tensor method --- has been adopted in shearing box
simulations \citep{jiang+12}. However, due to its non-local character,
implementing this scheme in GR is currently not feasible. 

A natural
closure scheme for global simulations in GR is 
M1 closure  \citep{levermore84} which evolves the radiative
flux independently of the radiative energy density. It has
been recently implemented into a GR hydrodynamical code
\koral\, \citep{sadowski+koral}, for neutrino radiation transport
in a GRMHD code \citep{shibata+12}, and into a special-relativistic magnetohydrodynamical (MHD)
code by \cite{takahashiohsuga-13}.

In this work we introduce an MHD version of the GR radiation code
\koral, verify the code by solving a number of test problems (including 
relativistic radiation-modified magnetosonic linear waves), and apply
it to a BH accretion problem by simulating super-critical  accretion  onto a BH.

The structure of the paper is as follows: In Section \ref{s.equations} we
introduce the equations describing the evolution of gas, magnetic field, and
radiation. In Section \ref{s.koral} we give details of the numerical code 
followed by a description of test problems in Section \ref{s.tests}.
In Section \ref{s.disks} we introduce and describe two simulations of super-critical
accretion we have performed. We conclude and discuss the implications in 
Section \ref{s.summary}.

\section{Equations}
\label{s.equations}

\subsection{Conservation of mass, energy and momentum}
\label{s.conservation}

The conservation laws for a fluid can be written in the covariant form,
\bea\label{eq.cont}
\hspace{1in}(\rho u^\mu)_{;\mu}&=&0,\\\label{eq.tmunucons}
\hspace{1in}(T^\mu_\nu)_{;\mu} &=& 0, 
\eea 
where $\rho$ is the gas
density in the comoving fluid frame, $u^\mu$ is the gas four-velocity
as measured in the ``lab frame'', and $T^\mu_\nu$ is the
MHD stress-energy tensor in this frame \citep{gammie03},
\be\label{eq.tmunu}
T^\mu_\nu = (\rho+u+p+b^2)u^\mu u_\nu + (p+\frac12b^2)\delta^\mu_\nu-b^\mu b_\nu.
\ee 
Here $u$ and $p=(\Gamma-1)u$ represent the internal energy and pressure of the 
gas in the comoving frame and $b^\mu$ is the magnetic field 4-vector \citep{gammie03}.

The conservation of energy and momentum of the radiative field can
be described by introducing the radiative stress-energy tensor $R^\mu_\nu$
\citep[][Section~\ref{s.rmunu}]{mihalasbook} which satisfies, assuming zero opacities for the moment, 
\be\label{eq.rmunucons}
(R^\mu_\nu)_{;\mu} = 0.  
\ee 
In more general case of non-zero opacities, the gas and radiation field
are coupled and must satisfy conservation of  total energy and momentum,
\be
\label{eq.totcons}
(T^\mu_\nu+ R^\mu_\nu)_{;\mu} = 0.  
\ee
The interaction between the gas and the radiation may be described with
the help of the radiation four-force density $G_\nu$ (Section~\ref{s.rmunu})
such that,
\bea\label{eq.cons2}
\hspace{1in}(T^\mu_\nu)_{;\mu}&=&G_\nu,\\\nonumber
\hspace{1in}(R^\mu_\nu)_{;\mu}&=&-G_\nu.
\eea
The opposite signs of the source terms on the right hand sides
of the two equations reflect the fact
that the gas-radiation interaction is conservative, i.e., it transfers energy
and momentum between gas and radiation.

The rest mass conservation equation~(\ref{eq.cont}) and the
energy-momentum conservation equations~(\ref{eq.cons2}) 
may be written in a coordinate basis in the 
following conservative form \citep{gammie03},
\bea\label{eq.cons3_1}
\hspace{.3in}\partial_t(\gdet\rho u^t)+\partial_i(\gdet\rho u^i)&=&0,\\\label{eq.cons3_2}
\hspace{.3in}\partial_t(\gdet T^t_\nu)+\partial_i(\gdet T^i_\nu)&=&\gdet T^\kappa_\lambda \Gamma^\lambda_{\,\,\nu\kappa} + \gdet G_\nu,\\\label{eq.cons3_3}
\hspace{.3in}\partial_t(\gdet R^t_\nu)+\partial_i(\gdet R^i_\nu)&=&\gdet R^\kappa_\lambda \Gamma^\lambda_{\,\,\nu\kappa} - \gdet G_\nu,
\eea
where $\gdet$ is the metric determinant, and $\Gamma^\lambda_{\,\,\nu\kappa}$ are the Christofel symbols.

\subsection{Maxwell's equations}

We adopt the ideal MHD approximation and assume that the electric field 
vanishes in the fluid rest frame. Maxwell's equations may be  
written  as \citep{gammie03},
\be
F^{*\mu\nu}_{;\nu}=0,
\ee
where $F^{*\mu\nu}=b^\mu u^\nu - b^\nu u^\mu$ is the Maxwell
electromagnetic tensor,
and $b^\mu$ is the four-vector of comoving magnetic field.
They can be further simplified by introducing the magnetic field three-vector $B^i=F^{*it}$
\citep{komissarov-99} which satisfies,
\be
\label{eq.Bit}
b^t=B^i u^\mu g_{i\mu},
\ee
\be
\label{eq.Bi1}
b^i=\frac{B^i+b^tu^i}{u^t}.
\ee
In terms of $B^i$, Maxwell's equations in the coordinate basis are,
\be
\label{eq.Maxi}
\partial_t(\sqrt{-g}B^i)=-\partial_j\left(\sqrt{-g}(b^ju^i-b^iu^j)\right),
\ee
\be
\label{eq.Maxt}
\frac1{\sqrt{-g}}\partial_i(\sqrt{-g}B^i)=0.
\ee
The first equation is the induction equation which describes the oveolution of the 
magnetic field. The second equation is the divergence-free condition and is not evolved directly. Instead, 
the flux-interpolated contrained transport (Flux-CT) method of \cite{toth-00} is used 
to prevent numerical generation of magnetic monopoles.

\subsection{Radiative stress-energy tensor and four-force}
\label{s.rmunu}

The radiation stress-energy tensor in an orthonormal frame
  comprises various moments of the specific intensity $I_\nu$,
  e.g., in the fluid frame it takes the following form,
\be  \label{eq.Rmunuff}
\widehat R= \left[ \begin{array}{cc} \widehat E & \widehat F^i \\ \widehat F^j &
    \widehat P^{ij}
\end{array} \right],
\ee
where the fluid-frame quantities, denoted with ``hats'',  are defined as,
\bea
\hspace{1in}\widehat E&=& \int \widehat I_\nu {\,\rm d \nu \,d\Omega},\\
\hspace{1in}\widehat F^i&=& \int \widehat I_\nu {\,\rm d \nu \,d\Omega \,N^i},\\
\hspace{1in}\widehat P^{ij}&=& \int \widehat I_\nu {\,\rm d \nu \,d\Omega \,N^i\,N^j}.
\eea
$\widehat E$ is the radiation energy density, $\widehat F^i$ is the radiation flux and 
$\widehat P^{ij}$ is radiation
pressure tensor, and $N^i$ is the unit vector in the direction of
$\partial/\partial x^i$.

In the same frame, the four-force $\widehat G^\nu$ is given by \citep{mihalasbook},
\be
\widehat G^\nu=\int(\chi_\nu I_\nu - \eta_\nu){\,\rm d \nu \,d\Omega \,N^i},
\ee
where $\chi_\nu$ and $\eta_\nu$ denote the frequency-dependent
opacity and emissivity coefficients.
$\widehat G^\nu$ takes a particularly simple form for grey opacities,
\be\label{eq.Gff}
\widehat G=
\left[ \begin{array}{c}
 \kappa_{\rm a}\rho (\widehat E-4\pi \widehat B)\\
 \chi\rho \widehat F^i 
\end{array} \right].
\ee Here, $\widehat B=\sigma T^4/\pi$ is the integrated Planck
function corresponding to the gas temperature $T$, $\sigma$ is the
Stefan-Boltzmann constant, $\kappa_{\rm a}$ and $\chi$ are
the grey absorption and total opacity coefficients, respectively. The
total opacity consists of the absorption ($\kappa_{\rm a}$) and scattering
($\kappa_{\rm es}$) opacities, i.e., $\chi=\kappa_{\rm a}+\kappa_{\rm es}$.

The comoving-frame radiative four force (Eq.~\ref{eq.Gff}) may be written
in the following form,
\be\label{eq.Gcov1}
\frac1\rho\left[ \begin{array}{c}
 \widehat G^t\\
 \widehat G^i
\end{array} \right]
 = (\kappa_{\rm a}+\kappa_{\rm es})
\left[ \begin{array}{c}
 \widehat E\\
 \widehat F^i
\end{array} \right] - 
\left[ \begin{array}{c}
 (\kappa_{\rm es} \widehat E + \kappa_{\rm a} 4\pi \widehat B)\\
 0
\end{array} \right].
\ee The contravariant and covariant orthonormal comoving-frame gas velocities are
$\widehat u^\mu=(1,0,0,0)$, and $\widehat u_\mu=(-1,0,0,0)$,
respectively. From Eq.~(\ref{eq.Rmunuff}) it follows that,
\be \label{eq.EFRmunu}
\widehat R^{\mu\nu} \widehat u_\nu = -\left[ \begin{array}{c}
 \widehat E\\
 \widehat F^i,
\end{array} \right] \ee
and
\be \label{eq.EhatRmunu}
\widehat R^{\mu\nu} \widehat u_\mu \widehat u_\nu = \widehat E.
 \ee
Using these definitions in Eq.~(\ref{eq.Gcov1}) we obtain,
\be
\label{eq.Gcov2}
\widehat G^\mu = -\rho (\kappa_{\rm a}+\kappa_{\rm es})\widehat R^{\mu\nu} \widehat u_\nu
-\rho \left(\kappa_{\rm es}\widehat R^{\alpha\beta} \widehat u_\alpha \widehat u_\beta+\kappa_{\rm a} 4\pi \widehat B\right)\widehat u^\mu.
\ee
This expression is written in a covariant form, so it is valid in any frame. Therefore,
we can skip the ``hats'' and obtain the following expression for the radiative four-force, which is valid in an arbitrary frame,
\be
\label{eq.Gcov}
 G^\mu = -\rho (\kappa_{\rm a}+\kappa_{\rm es}) R^{\mu\nu}  u_\nu
-\rho \left(\kappa_{\rm es} R^{\alpha\beta}  u_\alpha  u_\beta+\kappa_{\rm a} 4\pi  B\right) u^\mu.
\ee
Note that for non-trivial four-velocities $u^\mu$, the four-force depends
on all the components of $R^{\mu\nu}$, not just the lab frame energy density ($R^{tt}$)
and flux ($R^{ti}$).

\subsection{Closure scheme}
\label{s.closure}
To close the above set of equations we need a prescription to compute
the second moments of the radiation  intensity.
Specifically, we need a prescription to write down the full radiation
stress tensor $R^{\mu\nu}$ knowing only the radiative energy density
$R^{tt}$ and the fluxes $R^{ti}$. We follow the M1 scheme,
introduced by \cite{levermore84}, and assume that the radiation
tensor is isotropic, and satisfies the Eddington closure, not in the
fluid frame, but in the orthonormal ``rest frame'' of the
radiation. The latter is defined as the frame in which the radiative
flux vanishes. 

\cite{sadowski+koral} introduced a covariant formalism for the M1 scheme.
Herein, we give only the essential formulae and ask the reader to refer to
that paper for details.

Knowing $R^{t\mu}$ we may calculate the time-component of the radiative rest-frame
four velocity, $u^t_R$, and radiative energy density in this frame, $E_R$, by solving
a set of two equations,
\be
g_{\mu\nu}\,R^{t\mu}R^{t\nu} = -\frac{8}{9}E_R^2 (u^t_R)^2
+\frac{1}{9}E_R^2 g^{tt},
\ee
\be
R^{tt} = \frac{4}{3}E_R (u^t_R)^2 + \frac{1}{3}E_R g^{tt}.
\label{eq:invert2}
\ee
These quantities may be then used to find the spatial components of $u^\mu_R$
using the time component of,
\begin{equation} 
R^{\mu\nu} = \frac{4}{3}\bar E\, u^\mu_R u^\nu_R + \frac{1}{3}E_R\,
g^{\mu\nu}.
\label{eq:R}
\end{equation}
Once the four-velocity of the radiative rest frame, $u^\mu_R$, and the radiative
energy density in that frame, $E_R$, are known, one can use Eq.~(\ref{eq:R}) again
to find all the unknown components of the radiative stress energy tensor $R^{\mu\nu}$.

The M1 closure scheme handles well both extremes of the 
optical depth and it is found to be fairly good at
intermediate optical depths as well. However, the implicitly assumed specific intensity is always
symmetric with respect to the mean flux. The M1 closure scheme is thus
expected to be inaccurate when multiple sources of radiation are involved.
In problems involving accretion disks, which are the primary interest of this paper, 
anisotropic configurations
with multiple beams are not very common, as the light is expected
to be emitted into the optically thin region from a continuous and
smooth photosphere, and the M1 scheme is probably
adequate. In any case, M1 closure provides a significantly
superior treatment of radiation in the optically thin regions near and
above the disk photosphere, compared to the Eddington approximation or
flux-limited diffusion. It is, in the same time, much less precise than non-local 
methods, e.g., the Variable Eddington Tensor approach \citep{jiang+12}, which are, 
unfortunately, difficult to implement in general relativity.

\section{The GRRMHD code --- \koral}
\label{s.koral}
The numerical code \koral\, was developed initially without magnetic fields
as a general relativistic radiation hydrodynamics code \citep{sadowski+koral}.
In this paper we describe the extension of \koral\, to GRRMHD. The basic methods 
we use are the same as in the original code  except that the radiation
implicit solver  has been updated as described in Section~\ref{s.implicit}.
Below we give a general description of the code and refer the reader to \cite{sadowski+koral}
for details not given here.

\subsection{The algorithm}
\label{s.algorithm}

The code uses a finite difference scheme with 
linear slope-limited reconstruction and
the van-Leer's one-parameter family of minmod limiters with the dissipation parameter
$\theta_{\rm MINMOD}=1.5$ if not mentioned otherwise. The fluxes at the cell faces
are calculated using the Lax-Friedrichs method. The source terms are
applied at the cell centers and the time stepping is performed using
the second order Runge-Kutta method. 

A set of 13 equations is evolved. It comprises of rest mass conservation
(Eq.~\ref{eq.cons3_1}), four energy and momentum conservation equations for the
MHD fluid (Eq.~\ref{eq.cons3_2}), and  for the radiation field (Eq.~\ref{eq.cons3_3}),
three components of the induction equation (Eq.~\ref{eq.Maxi}), 
and the entropy evolution equation, 
$(S u^\mu)_{;\mu}=0$, which in the coordinate frame is,
\be
\label{eq.entrev}
\partial_t(\gdet S u^t)+\partial_i(\gdet S u^i)=0,
\ee
where,
\be \label{eq.entr}
S=\frac{\rho}{(\Gamma-1)}
  \log\left(\frac{p}{\rho^\Gamma}\right),  
\ee
is the gas entropy per unit volume.

The
vector of conserved quantities is (see Section~\ref{s.implementation})
\be 
\label{eq.consvec}
U = [\rho u^t,T^t_t+\rho u^t,T^t_i,S u^t,B^i,R^t_t,R^t_i], 
\ee 
while the
primitive quantities are, \be P = [\rho, u, u^i, S, B^i, E_R, u^i_R].  \ee
Conversion from conserved to primitive quantities is described in
Section~\ref{s.conversions}.

During each sub-step of the Runge-Kutta time integration, the code
carries out the following steps in the given order:
\begin{enumerate}
\item The conserved quantities in each cell are used to
  calculate the primitive quantities at the cell center (Section~\ref{s.conversions}). 
Floors (Section~\ref{s.implementation}) are imposed if necessary.
\item Ghost cells at the boundaries of the computational domain are
  assigned primitives appropriate to the boundary conditions of the
  particular problem of interest.
\item For each cell, the maximal characteristic left- and right-going
  wave speeds are calculated. The MHD wave speeds are calculated as given in 
\cite{gammie03}, while the radiative characteristic speed is calculated as 
in the original code \citep{sadowski+koral}.
\item For each dimension, primitives are interpolated 
to obtain their left- and right-biased values
  at cell faces: $P_L$ and $P_R$ which are verified against floors.
\item From $P_L$ and $P_R$, left- and right-biased fluxes ${\cal F}_L$
  and ${\cal F}_R$ are calculated at cell faces and are then combined using the Lax-Friedrichs scheme.
\item These combined fluxes are modified following the flux-interpolated constrained transport
algorithm \citep{toth-00} to preserve divergence-free magnetic field.
\item The advective time derivative is calculated using an unsplit
  scheme,
\be\label{eq.lax}
\frac {dU}{dt}_{\rm (adv)} = -\frac{{\cal F}^1_R-{\cal F}^1_L}{dx^1}-\frac{{\cal F}^2_R-{\cal F}^2_L}{dx^2}-\frac{{\cal F}^3_R-{\cal F}^3_L}{dx^3},
\ee
where $dx^i$ denotes cell size in the direction $\partial/\partial x^i$.
\item The geometrical source terms, i.e., the terms
involving Christofell symbols on the right hand side of Eqs.~(\ref{eq.cons3_2})
and (\ref{eq.cons3_3}) are calculated at cell centers to
  give the corresponding time derivative ${dU}/{dt_{\rm (geo)}}$.
\item The advective and geometrical operators are used to update the
  conserved quantities according to
\be
\Delta U = \left(\frac {dU}{dt}_{\rm (adv)}  + \frac {dU}{dt}_{\rm (geo)}  \right) \Delta t.
\ee
That is, all these terms are treated in an explicit fashion.
\item The updated vectors of conserved quantities are used to
  calculate the corresponding updated primitive quantities at cell
  centers. Floors are applied if required.
\item Finally, the remaining four-force density $G_\nu$, is handled implicitly using the method
  described in Section~\ref{s.implicit}. This results in a final
  update of the vector of conserved quantities at each cell center.
\end{enumerate}

\subsection{Conversion between conserved and primitive quantities}
\label{s.conversions}

Throughout the algorithm the vectors of conserved and primitive quantities have
to be converted one to the other multiple times.  In our problem, the
MHD and radiative variables decouple, so the conversion may
be done separately for each. 

In GR numerical codes starting from MHD primitive quantities and calculating the vector of MHD conserved quantities
is algebraic and straightforward.  The opposite 
calculation (inversion) requires a numerical solve for which we use the ``$1D_W$'' inversion
scheme described in \cite{nobleetal06}.

The radiative conserved variables ($R^t_\mu$) may  in principle be
easily converted to the radiative primitives ($E_R$, $u^i_R$)
following the algorithm described in Sect.~\ref{s.closure}.  However,
it requires solving a quadratic equation which may not provide a
finite and positive solution for $u^t_R$. Such a
case corresponds to unphysical flux, $\widehat F > c\widehat E$, where
$c$ stands for the speed of light, and may appear because of truncation error, especially near
radiation fronts in optically thin regions. To handle such failed
inversions we increase the lab-frame energy density, $R^{tt}$, to
achieve a prescribed limiting value of the radiation rest frame
Lorentz factor,
 $\gamma_{R,{\rm max}}$. Details of this 
procedure are given in the companion paper \cite{mckinney+harmrad}.

\subsection{Implicit treatment of radiative source terms}
\label{s.implicit}

The radiative source term $\pm G_\nu$ in Eqs.~(\ref{eq.cons3_2}) and
(\ref{eq.cons3_3}) becomes stiff in optically thick regime and has to
be handled implicitly \citep[e.g.,][]{zanottietal11}. In both the original
and the GRRMHD version of the \koral\, code we take the semi-implicit
approach, i.e., we split the advective operator (spatial derivatives
in Eqs.~\ref{eq.cons3_1}-\ref{eq.cons3_3})
from the radiative
source term operator (terms proportional to $G_\nu$) and apply the latter locally. Such an approach
is possible because the time step is already limited by the speed of
light just from the fluid dynamics, so advection of radiation is
guaranteed to be stable in an explicit scheme. The implicit solver
has been updated and is described in detail below.

The radiative source term operator describes the interchange of 
the energy and momentum density between the gas and radiation field.  
The corresponding equations are,
\bea
&&\hspace{2cm}\partial_t(T^t_\nu)=G_\nu,\label{eq.source1}\\
&&\hspace{2cm}\partial_t (R^t_\nu)= - G_\nu, \label{eq.source2}
\eea
which can be put in the implicit form as,
\bea
&&\hspace{1cm}T^t_{\nu,(n+1)}-T^t_{\nu,(n)}=\Delta t ~G_{\nu,(n+1)},\label{eq.source3}\\
&&\hspace{1cm}R^t_{\nu,(n+1)}-R^t_{\nu,(n)}= - \Delta t~ G_{\nu,(n+1)}, \label{eq.source4}
\eea
where the subscripts $(n)$ and $(n+1)$ denote values at the beginning and end of a 
time step of length $\Delta t$, respectively.

Because of the symmetry of the problem, specifically, the
right-hand sides of eqs.~(\ref{eq.source3})
and (\ref{eq.source4}) differing only by sign,
the system of equations may be reduced to four
non-linear equations. In the original version of \koral\, we
chose the radiative conserved quantities, $R^t_\mu$, as the
quantities we iterated inside the numerical solver. This
approach, however, turned out not to be very robust.
It  fails when the energy densities of the gas and radiation
are many orders of magnitude different, or when iteration steps
lead to unphysical states, preventing a proper radiative or MHD inversion.
The algorithm we propose here uses radiative or MHD primitives (in place
of conserved quantities) as the iterated quantities, prevents the iterations
from going out of bounds, and reverts to the entropy
conservation equations when the energy equation does not provide a solution. 
We solve the equations in the lab frame, but if that
fails, we try again with the energy equation written in the fluid-frame.
The scheme adopted here is not unique, 
and other approaches turn out to be efficient as well 
\citep[e.g.,][]{mckinney+harmrad}.

The semi-implicit scheme implemented in \koral\, is decribed in detail below.
We assume that it iterates the MHD subset of primitive quantities, though 
the algorithm is the same even for the case of radiative primitives.

\begin{enumerate}

\item The complete set of primitives (initial guess or current iteration stage), 
$P_{(n+1)}$, is used to calculate
the radiative four force $G_{\mu,(n+1)}$. 
\item Residuals $f^\mu$ and the error measure $\rm err$ are calculated through,
\be
\label{eq.resfmu}
f^\mu = T^t_{\mu,(n+1)} - T^t_{\mu,(n)}-\Delta t G_{\mu,(n+1)},
\ee
\be
{\rm err} = {\rm max}\left( \frac{|f^\mu|}{|T^t_{\mu,(n+1)}| + |T^t_{\mu,(n)}|+|\Delta t G_{\mu,(n+1)}|}\right),
\ee
where $T^t_{\mu,(n+1)}$ are the MHD conserved quantities obtained from $P_{(n+1)}$.
If ${\rm err}<10^{-8}$ the solution is found.

In case of the fluid-frame-based method, the time component of Eq.~\ref{eq.resfmu} 
is replaced by,
\be
\label{eq.resfmuff}
f^t = u_{(n+1)} - u_{(n)}-\Delta \tau \kappa_{\rm a} (\widehat E_{(n+1)} - 4\pi B_{(n+1)}),
\ee
with $\Delta \tau=\Delta t/u^t$. The error measure is calculated appropriately.
\item In a similar fashion we estimate the Jacobi matrix, 
\be
{\cal J_{\mu\nu}}=\frac{df^\mu}{dP_{\nu(n+1),\rm MHD}},
\ee by repeating the procedure for slightly (we choose factor of $10^{-8}$) modified
values of each of the MHD primitives. If such a sub-step leads to an unphysical state,
e.g., $\widehat F>c\widehat E$, we recalculate the residuals modifying the quantity
of interest by the same amount but with the opposite sign.
\item
The iterated primitives are then updated using the inverse of the Jacobi matrix, ${\cal J}^{-1}$
(this is the Newton-Raphson method),
\be
P'_{\mu,(n+1),\rm MHD}=P_{\mu,(n+1),\rm MHD} - \xi {\cal J}^{-1}_{\mu\nu} f^{\nu},
\ee
where the factor $\xi$ is initially set to $1$.
\item If the new time-component of the primitive vector, i.e., the energy density, is negative, the
$\xi$ factor is reduced appropriately (using linear interpolation) not to leave the
physical regime.
\item Knowing the new vector of MHD primitives, $P'_{(n+1),\rm MHD}$,
we calculate the corresponding set of conserved quantities, $U'_{(n+1),\rm MHD}$.
\item We use the conservation of energy and momentum
to calculate the radiative conserved quantities, $U'_{(n+1),\rm RAD}$, through,
\be
R^t_{\mu,(n+1)}=R^t_{\mu,(n)} - (T^t_{\mu,(n+1)}-T^t_{\mu,(n)}).
\ee
\item We perform the radiative inversion to obtain the corresponding set of
radiative primitives $P_{(n+1),\rm MHD}=\{E_{\rm R}, u^i_{\rm R}\}$. We note that this
is the only inversion that has to performed, in contrast to the method described
in \cite{sadowski+koral} where full (radiative and MHD) inversion was required.
Moreover, we stress that the radiative inversion is given in a closed form
and does not require numerical solvers. Therefore, iterating MHD
primitives is less computationaly expensive than iterating radiative primitives.
\item If the inversion fails (corrections were applied indicating an unphysical state), the factor $\xi$ is reduced by a factor of $2$ 
and $P'_{(n+1),\rm MHD}$ are recalculated (step iv). This procedure is repeated until a physical
state is found or until $\xi<10^{-10}$ in which case a failure is reported.
\end{enumerate}

Having in mind that the iterated quantites should be the sub-dominant ones 
(in the other case even small shifts in the dominant primitives
may have an overwhelming effect on the sub-dominant ones) we choose
to solve initially using the MHD primitives if
$u_{(n)}<100\widehat E_{(n)}$, i.e., we compare fluid frame energy densities
after the advection operator has been applied. If implicit solver fails, we
revert to the other set of primitives. If both methods fail (rare event, see below) 
we try them again in the same order but this time replacing the energy equation (Eq.~\ref{eq.resfmu}),
\be
f^t=T^t_{\mu,(n+1)} - T^t_{\mu,(n)}-\Delta t G_{\mu,(n+1)},
\ee
with the fluid-frame entropy evolution equation,
\be
f^t=S_{(n+1)} - S_{(n)} -\Delta \tau \kappa_{\rm a} (\widehat E_{(n+1)} - 4\pi B_{(n+1)}),
\ee
where $S$ is the entropy per unit volume given in Eq.~\ref{eq.entr}.

The method described above is robust and fails only in
pathological situations which may occur sporadically near shock fronts
or near the axis.
%To handle such cases we use a backup method described in
%Appendix~\ref{ap.backup} which calculates first the local thermal
%equilibrium (LTE) state and then interpolates the gas and radiation
%fields toward that state. If for some reason the backup method fails
%as well, 
In such a case 
the code averages out the problematic cell by taking the
average of primitives in the neighbouring cells which have evolved
properly.

In the developed stage of a disk-jet simulation with BH spin $a_*=0.9$
described in Section~\ref{s.disks} the implicit solver works in 97\%
of cases on the most-robust MHD primitives, and on radiative
primitives in the remaining 3\% of cases. The average numbers of
iterations within the solver are 2.2 and 3.1, respectively.  The solver
practically never reverts to the entropy evolution. The latter is used 
very sporadically in the initial stages of the simulation when the empty space outside the torus is abrubtly filled first
by the radiation fields, and then by magnetic fields. Once
this initial transient adjustment is over,
the energy evolution equation works 100\% of the time.

\subsection{Implementation notes}
\label{s.implementation}
\begin{enumerate}

\item 
The mass conservation (Eq.~\ref{eq.cons3_1}) and the gas internal
energy conservation law (the $t$ component of
Eq.~\ref{eq.cons3_2}), are aggregated to give,
\be
\partial_t\left(\sqrt{-g}(T^t_t+ \rho u^t)\right)+\partial_i\left(\sqrt{-g}(T^i_t+
\rho u^i)\right)=\sqrt{-g}T^\kappa_\lambda \Gamma^\lambda_{\,\, t\kappa} +
\sqrt{-g}G_t.
\label{eq.hdlab4}
\ee
%which replaces the $t$ component of equation~(\ref{eq.cons3_2}).
%  Then,
$T^t_t + \rho u^t$ becomes the relevant conserved quantity (Eq.~\ref{eq.consvec}), which
reduces in the non-relativistic limit to minus internal energy density
of gas.

\item In cold relativistic flows, where $u\ll \rho$, the numerical
  accuracy is not sufficient to evolve the internal energy reliably.
  As a result, negative internal energy densities may be occasionally
  found. Whenever this happens, we perform the MHD inversion using
the independently evolved value of entropy, $S u^t$, instead of $T^t_t$.
In the case of successful energy-based inversion, we update the
value of entropy according to the new gas primitives and evolve
it independently in the next time step.

%\item At the beginning of each Runge-Kutta sub-step we
%compare the value of the independently evolved entropy with the
%entropy corresponding to the density and pressure obtained by
%the energy-based inversion. The entropy increase principle requires
%the latter to be no smaller than the former.

\item
Because of the truncation error, the geometrical source terms
involving Christoffel symbols (Eqs.~\ref{eq.cons3_2} and \ref{eq.cons3_3}) 
do not balance the corresponding spatial derivatives on the
left hand side 
even for particularly simple situations such as constant gas or
radiation pressure, and can lead to catastrophic secular errors. To
solve this issue we modify the values of the Christoffel symbols
suitably following Appendix A of \cite{mtb12}.

\item
The vector of primitives
includes spatial components of gas and radiation rest frame four-velocities $u^i$ and $u^i_{\rm R}$. 
Inside the BH ergosphere they do not determine corresponding values of $u^t$ uniquely (two 
time-like observers may have the same spatial components of four-velocity).
Therefore, instead of the regular four-velocity, $u^i$, we use
velocity $\tilde u^i$ defined with respect to the normal 
(with four velocity $\eta_t=\sqrt{-1/g^{tt}},\,\eta_i=0$) observer. The conversions
between the velocities read,
\be
\tilde u^i=u^i - u^t \frac{g^{ti}}{g^{tt}},
\ee
\be
u_\mu=\tilde u_\mu - \sqrt{\alpha^2 \gamma^2}\delta^t_{\mu},
\ee
where $\tilde u^t=0$, $\alpha^2=-1/g^{tt}$, $\gamma^2=1+\tilde u^i \tilde u_i$, and
$\delta^t_{\mu}$ is the Kronecker delta.

\item
To ensure stability of the code and to prevent unphysical runaway of mass from the polar
region in disk-jet simulations we impose a number of constraints ('floors' and 'ceilings') on the primitive
quantities. Firstly, we do not allow the internal energy density either to drop below 
$10^{-10}$ of the rest mass density, $\rho$, or to exceed it by a factor of $100$.
We impose a maximal Lorentz factor $\gamma=50$ on gas and the radiation rest frame.
If the magnetic pressure, $b^2/2$, exceeds $\rho$ by a factor of $100$ or more, we
introduce extra mass in the ZAMO frame and modify the internal energy density
to keep the temperature fixed. If $\rho u^t$ turns out to be negative, we average 
such a cell out using primitives from successfully evolved neighbors.

\item Units --- Radiative code requires proper coupling of radiation and gas physical
constants. Besides setting $G=c=1$, we choose to keep the kelvin as the unit of temperature. Table \ref{t.units}
gives code values of constants and the conversion factors between the code and cgs units.

\begin{table}
\begin{minipage}{.9\columnwidth}
\caption{Physical constants and conversion factors between code and cgs units. $M$ stands for the BH mass in cgs units.}
\label{t.units}
\begin{tabular}{@{}lc}
\hline
\hline
constant&code value\\
\hline
Boltzmann constant&$k_B=1.381\times 10^{-16}\, M^{-1}c^{-2}$\\
Stefan-Boltzmann constant&$\sigma = 5.670\times 10^{-5}\, G^3 M^2 c^{-9}$\\
Proton mass&$m_P=1.673\times 10^{-24}\, M$\\
\hline
\hline
quantity&code to cgs units conversion factor\\
\hline
length&$GM c^{-2}$\\
time  &$GM c^{-3}$\\
velocity & $c$\\
temperature & $1$\\
mass density & $c^6G^{-3}M^{-2}$ \\
energy density & $c^8G^{-3}M^{-2}$ \\
energy flux & $c^9G^{-3}M^{-2}$ \\
opacity & $G^2Mc^{-4}$\\
\hline
\end{tabular}
\end{minipage}
\end{table}

\end{enumerate}

\section{Test problems}
\label{s.tests}

The original \koral\, code was extensively tested in \cite{sadowski+koral}. The 
set of test problems solved there verified the evolution of relativistic hydrodynamical
fluids, optically thin radiation in curved spacetimes, and the interaction between
 gas and radiation. The GRRMHD version of \koral\, introduced in this paper solves properly all these previous tests.
In this Section we test the evolution of magnetic fields, which are a new feature in the code, and the coupling to gas and radiation. We start with a set of linear radiation-MHD
wave problems followed by non-linear tests: magnetic shock tubes and the Orszag-Tang vortex test problem.

\subsection{Radiation modified MHD linear waves}

We begin with studying propagation of linear waves  to test
accuracy of the coupled evolution of gas and radiation, and the
convergence rate of the code. We follow \cite{jiang+12} but
we recalculate the dispersion relation for a
relativistic fluid by perturbing Eqs.~(\ref{eq.cons3_1})-(\ref{eq.cons3_3})\footnote{
\texttt{Mathematica} notebook used to derive the dispersion
relation and solve for the eigenmodes is available at 
\texttt{http://cfa.harvard.edu/\textasciitilde asadowski/research/koral/rmhd.nb}}.
We set
a uniform background state on top of which we impose
perturbations corresponding to a given eigenmode and evolve
the system until the wave crosses the domain.  We studied a set of 9 linear waves
solved on a one dimensional grid. All three
spatial components of velocity, magnetic field, and radiation flux
were evolved. The wavelength was chosen to match the length of 
the domain spanning $0<x<1$. The absorption opacity and the velocity 
of the background state were set to zero, 
the internal energy density corresponded to the speed of sound
$c_s=0.1$, and the magnetic field to the Alvfen velocity
$v_A=0.2$. 

The parameters of
the background state and the perturbations are given in 
Table~\ref{t.waves}.
The top row corresponds to (left to right)
sonic, fast-magnetosonic and slow-magnetosonic waves for a fluid decoupled
from radiation. The second row gives the same modes but for
optically thin gas ($\tau=\kappa_{\rm es}=0.1$). The bottom row
corresponds to optically thick medium ($\tau=10$). For each
of the waves, $\mathbb{R}(\omega)/2\pi$ is the wave speed,
and $\mathbb{I}(\omega)$ determines the rate of damping 
resulting from transfer of energy between gas and radiation \citep{mihalasbook}.
All the non-radiative waves (top row) are not dissipative ($\mathbb{I}(\omega)=0$).

\begin{figure}
\includegraphics[height=.94\columnwidth,angle=270]{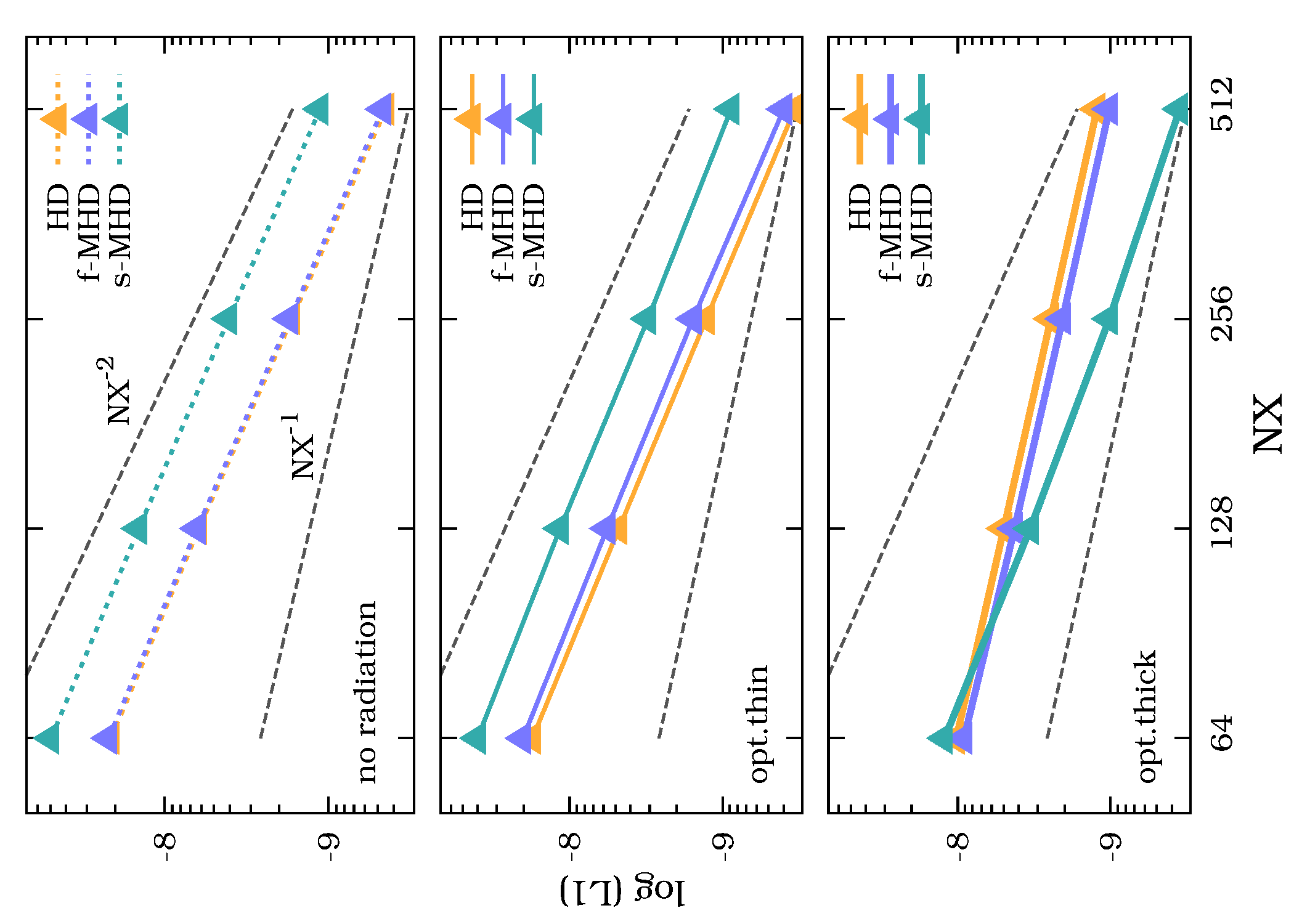}
\caption{Spatial convergence rate of the radiation modified MHD linear waves. The top panel shows 
pure (no radiation coupling) sonic (orange), fast-magnetosonic (blue), and slow-magnetosonic
(green) waves. The middle and bottom panels show the convergence rates of the corresponding
waves for optically thin ($\tau=0.1$) and thick ($\tau=10$) gas, respectively. The black dashed
lines show quadratic and linear convergence.}
  \label{f.rmhdl1}
\end{figure}

After time $t=2\pi/\mathbb{R}(\omega)$ we calculate the L1
error,
\be
L_1=\frac1{NX}\Sigma_{i=0}^{NX}\left(\rho_{\rm num}-\rho_{\rm ana}\right),
\ee
where $NX$ is the resolution, and $\rho_{\rm num}$ and $\rho_{\rm ana}$ correspond
to the numerical and analytical solutions, respectively. Fig.~\ref{f.rmhdl1}
presents the errors as a function of resolution. The top panel corresponds
to pure HD and MHD waves. They all converge as second order, as expected.
The middle panel shows the L1 error for optically thin waves. The rate of convergence
is the same because the evolution is dominated by the advective operator
and the radiative source term (which is applied implicitly) is insignificant.
It is not the case for optically thick gas. The rate of convergence 
is expected to be only linear because of the source term which dominates the evolution.
As the bottom panel shows, the radiation-modified sonic and fast-magnetosonic 
waves indeed show only linear convergence. The slow-magnetosonic wave
converges at an intermediate rate. 

\begin{table*}
\begin{minipage}{1.7\columnwidth}
\caption{Eigenmodes of linear waves. Other parameters: $\rho_0=1$, $u_0=9.13706\times 10^{-3}$, $u^x_0=u^y_0=0$, $\widehat F^x_0=\widehat F^y_0=0$,
$B^x_0=B^y_0=0.100759$, $\kappa_{\rm a}=0$, $k=2\pi$, $\Gamma=5/3$. Perturbations are of the form 
$\rho=\rho_0+\delta \rho\, e^{i(\omega t-k x)}$.
}
\label{t.waves}
\begin{tabular}{@{}lccc}
\hline
\hline
No.                    & $\tau=0$, $B=0$    & $\tau=0$, fast              & $\tau=0$, slow  \\
\hline 
$\delta\rho$           & $10^{-6}+0i$                                   & $10^{-6}+0i$                                   & $10^{-6}+0i$ \\
$\delta u$             & $1.52284\times 10^{-8}+0i$   & $1.52284\times 10^{-8}+0i$& $1.52284\times 10^{-8}+0i$\\
$\delta u^x$           & $1.0\times 10^{-7}+0i$ & $1.60294\times 10^{-7} + 0i$ & $6.17707\times 10^{-8}+0i$\\
$\delta u^y$           & $0+0i$ & $-9.79087\times 10^{-8} + 0i$ & $1.00118\times 10^{-7}+0i$\\
$\delta B^y$           & $0+0i$ & $1.62303\times 10^{-7}+0i$ &$-6.25516\times 10^{-8}+0i$ \\
$\omega$               & $0.628319+0i$                         & $1.00716+0i$                         & $0.388117+0i$ \\
\hline
\hline
No.                    & $\tau=0.1$, ${\cal P}=0.1, B=0$                & $\tau=0.1$, ${\cal P}=0.1$, fast & $\tau=0.1$, ${\cal P}=0.1$, slow \\
\hline 
$\delta\rho$           & $10^{-6}+0i$                                                                  & $10^{-6}+0i$ & $10^{-6}+0i$  \\
$\delta u$             & $1.51557\times 10^{-8}+7.69693\times 10^{-10}i$     & $1.51984\times 10^{-8}+4.81575\times 10^{-10}i$ & $1.50174\times 10^{-8}+1.22299\times 10^{-9}i$ \\
$\delta u^x$           & $9.97992\times 10^{-8}+2.55207\times 10^{-9}i$  & $1.60251\times 10^{-7}+7.23831\times 10^{-10}i$ & $6.15333\times 10^{-8}+1.83144\times 10^{-9}i$ \\
$\delta u^y$           & $ 0+0i$                                     & $-9.79544\times 10^{-8}+9.83679\times 10^{-10}i$ & $ 9.89772\times 10^{-8}+6.54186\times 10^{-9}i$ \\
$\delta B^y$           & $0+0i$                                  & $ 1.62344\times 10^{-7}-8.96662\times 10^{-10}i$& $-6.14882\times 10^{-8}-5.88315\times 10^{-9}i$\\
$\delta \widehat E$    & $1.33148\times 10^{-13}+3.60017\times 10^{-11}i$   & $1.48421\times 10^{-12}+6.06322\times 10^{-11}i$ & $1.91703\times 10^{-13}+2.18721\times 10^{-11}i$\\
$\delta \widehat F^x$  & $-2.52471\times 10^{-10}+7.40041\times 10^{-11}i$  & $-3.95433\times 10^{-10}+8.51051\times 10^{-11}i$& $-1.65181\times 10^{-12}+7.17520\times 10^{-11}i$\\
$\delta \widehat F^y$  & $0+0i$                                           & $2.36680\times 10^{-10}+2.11182\times 10^{-11}i$ & $-2.23679\times 10^{-10}-7.43141\times 10^{-11}i$\\
$\omega$               & $0.627057+0.0160351i$                            & $1.00689+0.00454797i$ & $0.386625+0.011507i$  \\
\hline
\hline
No.                    & $\tau=10$, ${\cal P}=10, B=0$    & $\tau=10$, ${\cal P}=10$, fast    & $\tau=10$, ${\cal P}=10$, slow  \\
\hline 
$\delta\rho$           & $10^{-6}+0i$                     & $10^{-6}+0i$                                                                    & $10^{-6}+0i$ \\
$\delta u$             & $1.17070\times 10^{-8}+1.88153\times 10^{-9}i$ & $1.17305\times 10^{-8}+1.71290\times 10^{-9}i$    & $9.46189\times 10^{-9}+1.21376\times 10^{-9}i$ \\
$\delta u^x$           & $2.66251\times 10^{-7}+6.33514\times 10^{-8}i$ & $2.78499\times 10^{-7}+5.23804\times 10^{-8}i$  & $8.34269\times 10^{-8}+1.20829\times 10^{-8}i$ \\
$\delta u^y$           & $0+0i$        & $-2.81093\times 10^{-8}+6.25588\times 10^{-9}i$ & $1.13633\times 10^{-7}+2.72697\times 10^{-7}i$ \\
$\delta B^y$           & $ 0+0i$           & $ 1.10170\times 10^{-7}-4.03337\times 10^{-9}i$  & $-8.03823\times 10^{-8}+3.03114\times 10^{-7}i$\\
$\delta \widehat E$    & $2.05419\times 10^{-7}+1.49859\times 10^{-7}i$& $2.07294\times 10^{-7}+1.36364\times 10^{-7}i$   & $2.59666\times 10^{-8}+9.67891\times 10^{-8}i$ \\
$\delta \widehat F^x$  & $-2.07308\times 10^{-8}+3.77556\times 10^{-8}i$& $-1.83331\times 10^{-8}+3.63664\times 10^{-8}i$  & $-1.98263\times 10^{-8}+5.47610\times 10^{-9}i$\\
$\delta \widehat F^y$  & $0+0i$    & $ 2.67581\times 10^{-10}+1.24272\times 10^{-9}i$ & $3.66075\times 10^{-9}-1.14750\times 10^{-9}i$ \\
$\omega$               & $1.6729+0.398049i$     & $1.74986+0.329116i$                                                & $0.524187+0.075919i$   \\
\hline
\hline
\end{tabular}
\end{minipage}
\end{table*}

\subsection{Magnetic shock tubes}

We choose two one-dimensional shock tube problems (2a and 5a) from
\cite{ryujones-95} to verify the non-linear evolution of a magnetized
fluid. We initialize the problem as left and right states separated
with a membrane which is removed at $t=0$. Table~\ref{t.magtubes}
lists gas parameters of both states. We have scaled the values from
\cite{ryujones-95} down to make the problem compatible with a
relativistic code (the corresponding velocity scaling factor, ${\cal
  C}$, is given in the table).  We solved the problem on a grid of 512
uniformly spaced points with $\theta_{\rm MINMOD}=2$.

\begin{table*}
\begin{minipage}{2.0\columnwidth}
\caption{Magnetic shock tubes}
\label{t.magtubes}
\begin{tabular}{@{}lccccccccccccccc}
\hline
\hline
Test & & & & \multicolumn{4}{c}{Left state:}&  &\multicolumn{4}{c}{Right state:} \\
No. & $\Gamma$ & ${\cal C}$ & & $\rho$ & $p$ & $u^i$ & $B^i$ &  &$\rho$ & $p$ & $u^i$ & $B^i$\\
\hline
2a & 5/3 & 100 & & 1.08 & $0.95/{\cal C}^2$ & $\left\{1.2,0.01,0.5\right\}/\,{\cal C}$ & $ \left\{2,3.6,2\right\}/\,4\pi {\cal C}$ 
 & & 1 & $1/{\cal C}^2$ & $\left\{0,0,0\right\}/\,{\cal C}$ & $ \left\{2,4,4\right\}/\,4\pi {\cal C}$ \\
5a & 5/3 & 100 & & 1 & $1/{\cal C}^2$ & $\left\{0,0,0\right\}/\,{\cal C}$ & $ \left\{0.75,1,0\right\}/\,4\pi {\cal C}$ 
 & & 0.125 & $0.1/{\cal C}^2$ & $\left\{0,0,0\right\}/\,{\cal C}$ & $ \left\{0.75,-1,0\right\}/\,4\pi {\cal C}$ \\
\hline
\hline
\end{tabular}
\end{minipage}
\end{table*}

\begin{figure}
\includegraphics[height=1.01\columnwidth,angle=270]{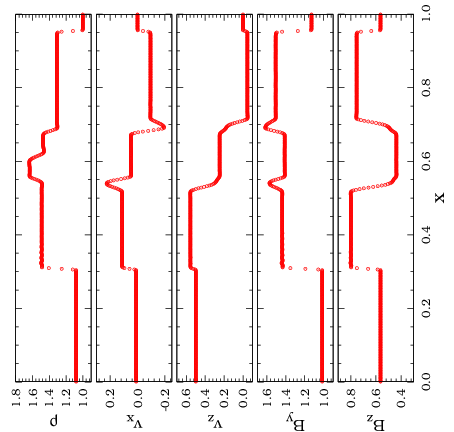}
\caption{Magnetic shock tube solution at $t=10$ for problem No. 2a.}
  \label{f.magtube2a}
\end{figure}

\begin{figure}
\includegraphics[height=1.01\columnwidth,angle=270]{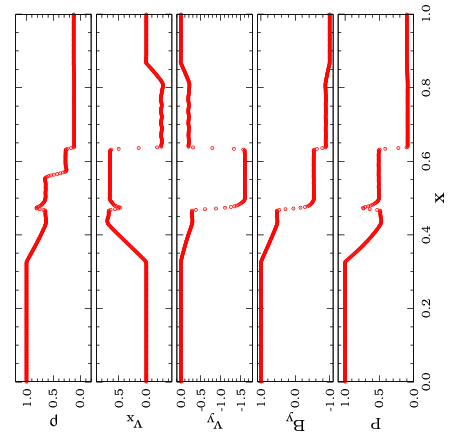}
\caption{Magnetic shock tube solution at $t=10$ for problem No. 5a.}
  \label{f.magtube5a}
\end{figure}

Figs.~\ref{f.magtube2a} and \ref{f.magtube5a} show the solutions for
problems 2a and 5a, respectively, at time $t=10$.  Profiles of (top to
bottom) density, two components of velocity, $y$-component of magnetic
field, and $z$-component of magnetic field (Fig.~\ref{f.magtube2a}) or
gas pressure (Fig.~\ref{f.magtube5a}) are shown. All quantities have been
scaled back to make the plots directly comparable with Figs. 6 and 7
from \cite{gammie03}. Similar plots may be found, e.g., in
\cite{ryujones-95} and \cite{stone+athena}. The agreement with these previous 
studies is very good.

\subsection{Orszag-Tang vortex}

The magnetic shock tubes tested one-dimensional evolution of magnetized
fluid. In this Section we perform a two-dimensional test which in addition
allows us to test the flux-constrained transport algorithm (the divergence is
trivially conserved in one-dimensional tests). For this purpuse we use the
standard Orszag-Tang vortex test problem \citep{orszagtang}. 

The initial state is set up
following \cite{stone+athena} but  rescaling all quantities by a velocity
factor ${\cal C}=100$. The density and pressure are initially uniform and set to
$25/(36\pi)$, and $5/(12\pi{\cal C}^2)$, respectively. The velocity field is  
$u^i=\left\{-\sin(2\pi y),\sin(2\pi x),0\right\}/\,{\cal C}$. The magnetic field
is set based on the vector potential  
$A^z=(B_0/4\pi)\cos(4\pi x) + (B_0/2\pi)\cos(2\pi y)$ with $B_0=1/\sqrt{4\pi}/{\cal C}$.
The problem is integrated over the domain $0<x,y<1$ with periodic boundary conditions, 
and 640x640 cells.

\begin{figure}

\includegraphics[width=1.075\columnwidth]{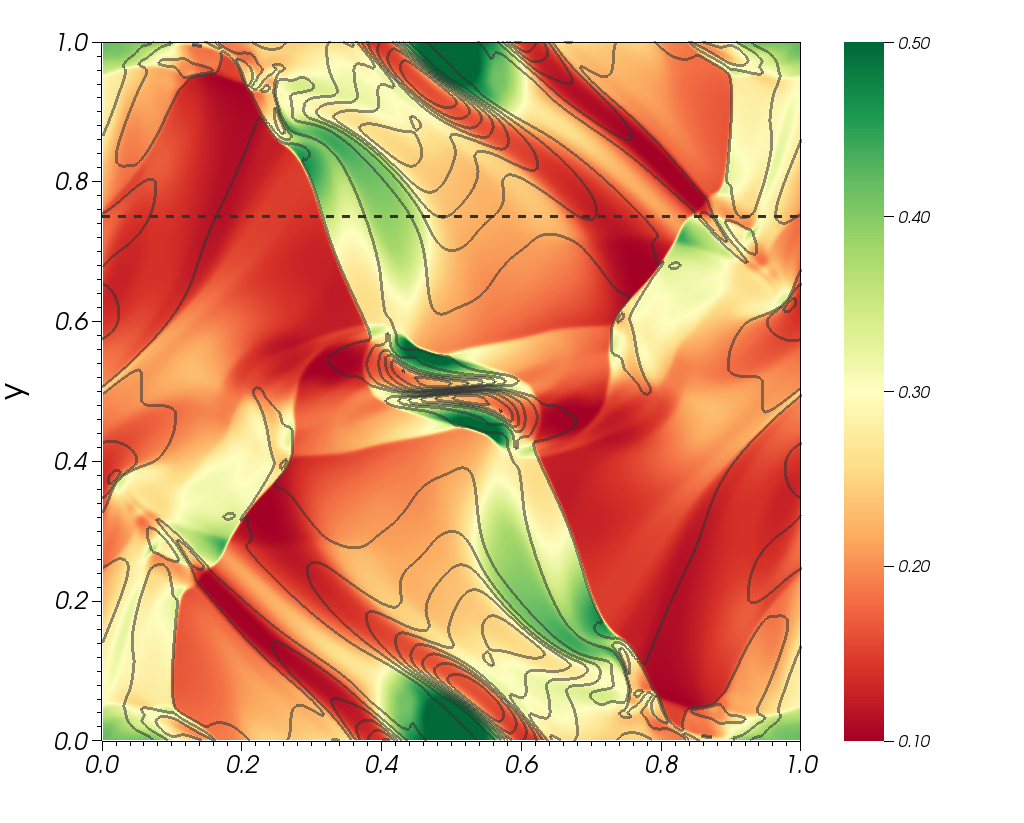}
\includegraphics[width=.9\columnwidth]{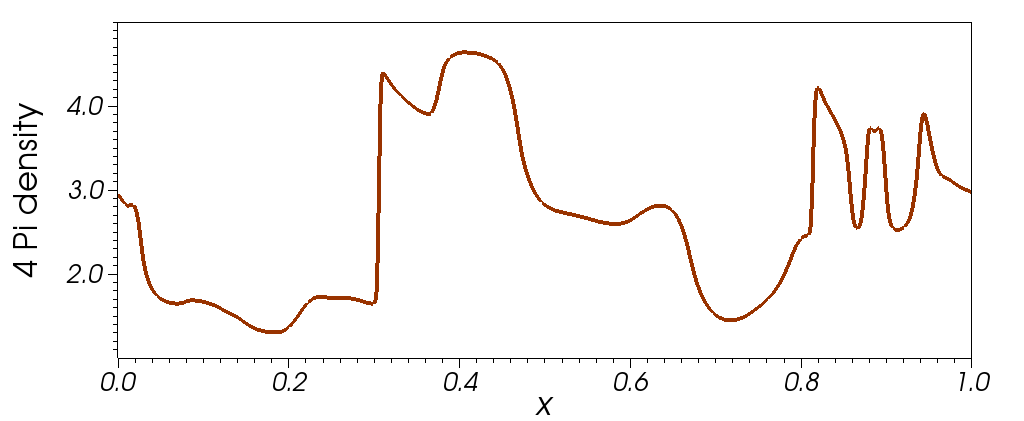}
\caption{Orszag-Tang test at $t=50.0$ solved with resolution 640x640. The top panel presents with colors the density distribution. Contours show distribution of magnetic pressure. The horizontal dashed line at $y=0.75$ corresponds to the slice shown in the bottom panel where density multiplied by $4\pi$ (for correspondence with \citet{gammie03}) is plotted.}
  \label{f.orszag}
\end{figure}

The top panel of Fig.~\ref{f.orszag} shows the density (colors) and magnetic pressure
(contours) at time $t=50$. These can be directly compared to Fig.~22 from 
\cite{stone+athena} (although their solution was calculated on a coarser grid
of 192x192 points). The bottom panel shows a slice in density along $y=0.75$ (denoted
by the dashed horizontal line in the top panel) which can be compared with
Fig.~9 of \cite{gammie03}. Again the agreement is very good.
We have also verified that $\partial_iB^i=0$ (Eq.~\ref{eq.Maxt}) throughout the evolution.

\section{Super-critical accretion}
\label{s.disks}

As a test application of the code we have simulated super-critical,
i.e., exceeding the Eddington limit, accretion on a non-rotating
($a_*=0.0$) and a spinning ($a_*=0.9$) BH. 

\subsection{Initial configuration}

\begin{figure*}
\includegraphics[width=2.13\columnwidth]{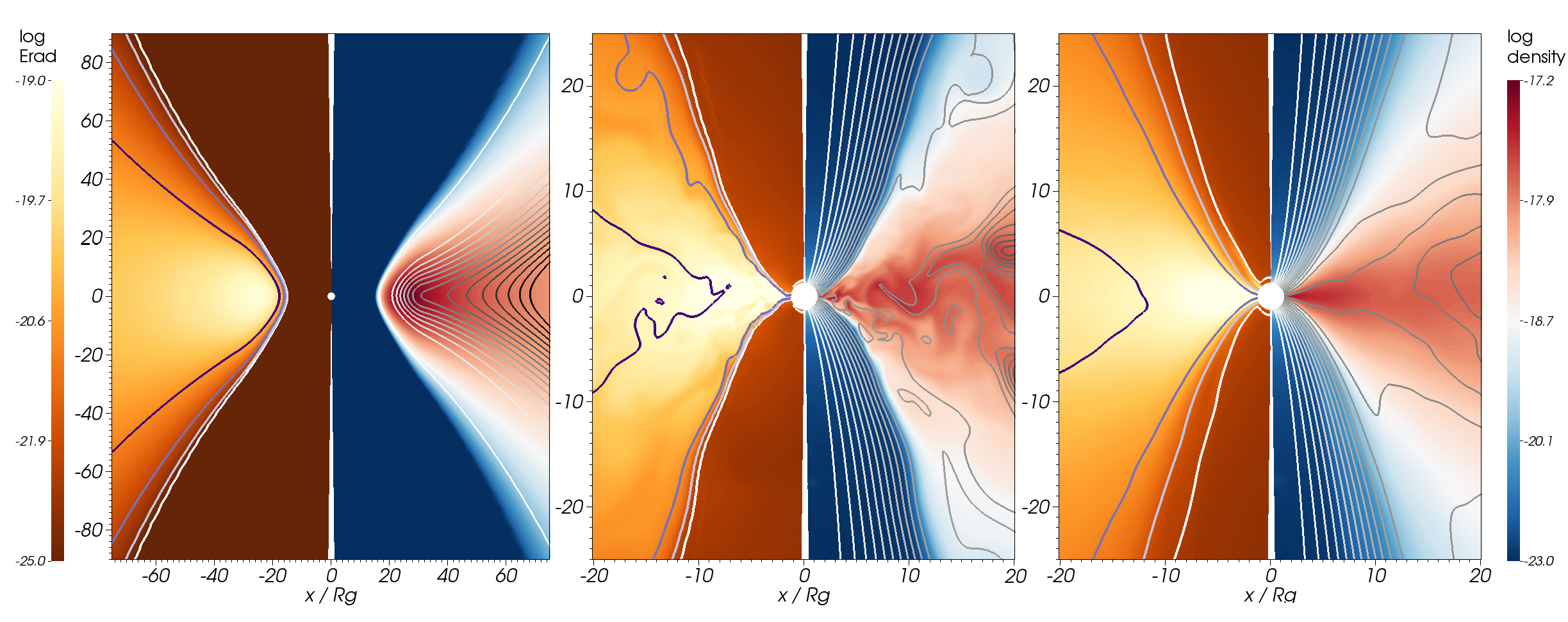}
\caption{Initial (left), snapshot at $t=1000$ (middle), and averaged
  over $t=2000\div12000\,GM/c^3$ (right panel) disk structure
in the $a_*=0.9$ simulation. In each panel, the left half shows the radiative energy
density in the comoving frame (colors) and the total optical depth
(contours), and the right half shows the density (colors) and magnetic
field lines (contours).}
  \label{f.400a9_triple}
\end{figure*}

 The left panel of
Fig.~\ref{f.400a9_triple} shows the initial state with the left half
showing the comoving frame radiative energy density ($\widehat E$),
and the right half showing the gas density. The equlibrium torus was
set up following \cite{penna-limotorus} with the parameters listed in
Table~\ref{t.limo}. The torus model provides the distribution of total
pressure, $p_{\rm tot}$, and gas density, $\rho$, assuming
$\Gamma=4/3$ (which is the correct value for a radiation pressure
dominated disk). This pressure is then distributed
between gas and radiation so as to satisfy  local thermal equilibrium
(LTE, $\widehat E=4\sigma T^4$). This involves solving the following quartic equation
for gas (and radiation) temperature, \be p_{\rm tot}=p_{\rm
  gas}+p_{\rm rad}=k_{\rm B}\rho T + \frac 43 \sigma T^4.  \ee 
Once the initial state has been set the polytropic index of the non-relativisitic 
gas is fixed at $\Gamma=5/3$ during the simulation. Because of the slight 
inconsistency in the value of $\Gamma$, the torus is not in the
perfect equilibrium at early times. However, it remains in a relatively steady state
until the magnetorotational instability (MRI) grows, after which the initial state is irrelevant.

The torus is initially threaded by a single poloidal loop of magnetic
field (left panel of Fig.~\ref{f.400a9_triple}) corresponding to a 
vector potential,
$$A_\phi = {\rm max}(R^4 \rho^2 10^{40} - 0.02, 0) \sin^4{\theta}.$$
The magnetic field is then normalized so that the plasma
parameter 
$\beta = p_{\rm tot} / p_{\rm mag}$, where $p_{\rm tot}$ and
$p_{\rm mag}$ stand for the total and magnetic pressure, respectively,
does not drop below $100$ at the equatorial plane.
The typical value
of the initial
$\beta$ near the torus pressure maximum ($R\approx 26$) is $\beta=400$.

The torus is surrounded by a low density and optically thin
atmosphere.  The velocities of both the gas and radiation rest frame
in the atmosphere correspond to the normal observer (relative
velocity $\tilde u^i=0$). The density and internal energy here are
set following $\rho = \rho_0 (R/2)^{-1.5}$, and $u=u_0 (R/2)^{-2.5}$, respectively,
with $\rho_0=10^{-24}$, and $u_0$ corresponding to temperature $10^{10}\rm K$. 
The radiative energy density, $E_R$, is set to a residual value of
$10^{-26}$ in the code units.

\begin{table}
\centering
\begin{minipage}{.7\columnwidth}
\caption{Equilibrium torus parameters \citep{penna-limotorus}}
\label{t.limo}
\begin{tabular}{@{}cccccc}
\hline
\hline
$R_{\rm in}$ & ${R_1}$ & $R_2$ & ${\xi}$ & $\kappa$ & $\Gamma$ \\
\hline
 15 & 31.75 & 200 & 0.9 & $1.5\times 10^{3}$ & $\frac43$ \\
\hline
\hline
\end{tabular}
\end{minipage}
\end{table}

\subsection{Numerical setup}

We evolved the disks in two dimensions, assuming axisymmetry, 
on a uniform grid of 200 points in the polar angle $\theta$ (spanning
$0.005\le \theta \le \pi-0.005$), and non-uniform, exponential grid of 300
points in radius. The internal ($x_1$) and Boyer-Lindquist ($R$) 
radial coordinates were related by,
$$R=e^{x_1},$$
and the range of uniform internal coordinates was chosen to correspond to
$1.75\le R\le500$ and $1.275\le R\le500$ for $a_*=0.0$ and $a_*=0.9$, respectively.
Such a choice of the inner radius ensures that the innermost 6 cells are 
located inside the BH horizon. We solved the problem on the appropriately
modified Kerr-Schild coordinates.

At the inner radial boundary we applied 
the outflow boundary condition by copying values of all the primitive 
quantities to the ghost cells. At the outer radial boundary we adopted
a similar approach, except that
we prevented the inflow of gas or radiation by resetting
negative radial velocities to zero. Reflective boundary condition
was adopted at the polar axis. To ensure stability in this region
the gas velocity in the two closest cells to the polar axis was
appropriately overwritten with the velocities from the third
cell \citep{mtb12}. Radial and azimuthal components were copied, and the polar
component was interpolated towards zero to satisfy the reflective
boundary condition.

To verify that the adopted resolution is enough to resolve
the fastest growing mode of the MRI we calculate the parameter
$Q_\theta$ \citep{hawley+11},
\be
Q_{\theta}=\frac{2\pi}{\Omega \Delta x^\theta}\frac{|B^\theta|}{\sqrt{\rho}},
\ee
where $\Delta x^\theta$ is the grid cell size in $\theta$. 
For the initial state of 
our simulations $Q_\theta\approx 10$ near the torus pressure maximum,
which is adequate.

\subsection{General properties}

At the onset of the simulation, some radiation (which fills
the entire initial torus) diffuses out through the torus surface and propagates
into the optically thin atmosphere. This initial flux 
results from the flux providing pressure support inside the torus.
Radiation propagates with the speed of light and initially
fills the entire space around the torus. After a while, the
angular component of the radiative flux, corresponding to
the rotation of the optically thick gas in the torus, prevents
radiation from reaching the polar axis and a temporary ``radiation 
shock'' forms close to the axis. This effect, resulting from
improper handling of collimating beams by the M1 scheme, is
discussed in more detail in Appendix~\ref{ap.a}. Once the disk
gets thicker and sub-Keplerian, radiation penetrates the entire polar
region and the shock dissapears.

Rotational shear makes the initial magnetic field unstable against
MRI. After a few orbital periods of the torus inner edge, the gas
becomes turbulent, loses angular momentum, and accretion begins. 
The middle panel of Fig.~\ref{f.400a9_triple} shows a snapshot of
the $a_*=0.9$ simulation taken after $t=10000$ which shows the
turbulent structure of the flow. The left and right halves of the
panel show comoving frame radiation and rest mass energy densities,
respectively. 

The contours in the left half show the total optical
depth (integrated from the polar axis) and correspond to $\tau_{\rm
  tot}=1$ (white), $10$, $100$, and $1000$ (deep blue).  The disk is
clearly very geometrically thick, as expected for a radiativelly
inefficient accretion flow. For cylindrical radius $R_{\rm cyl}=10$, 
the photosphere, defined as the surface where total optical depth 
measured from the polar axis is equal to unity, 
\be
\tau=\int_0^\theta \gamma \chi \sqrt{g_{\theta\theta}}\, \rm d\theta'=1,
\ee
 is located at angle
$\theta \approx 30^\circ$. The Lorentz factor $\gamma=u^t$ accounts
for the fact that the opacity $\chi$ is given in the comoving frame, and
we assume that the gas in the optically thin region moves 
radially, i.e., perpendicularly to path of integration \citep{rybicki-book}.
In contrast to the standard thin disk, the accretion flow 
does not terminate at the innermost stable circular orbit
(ISCO). This is consistent
with models of thick accretion disks \citep[e.g.,][]{abramowicz-isco}.

The contours in the right half show the lines of the poloidal magnetic field.
Initially, the field forms a single loop and is confined
to the equilibrium torus. Once the accretion starts, magnetic field is dragged
with the gas and deposited on the BH,  as reflected in 
the region of laminar, dense magnetic field lines in
the low-density funnel near the polar axis. The increased magnetic 
pressure in this region pushes on the disk from top and bottom and 
explains the decreased disk thickness in this region. Colors in the right half
of the middle panel show the distribution of density. It is strictly 
correlated to the distribution of comoving radiation energy density.
This is because inside the photosphere the scattering opacity is large and 
photons are trapped in the gas and radiation and gas move together.

The right panel of Fig.~\ref{f.400a9_triple} shows the time averaged
structure of the $a_*=0.9$ disk. All the turbulence has cancelled out
and the density and radiative energy density distributions are
smooth. 

\begin{figure*}
\includegraphics[width=1.631\columnwidth]{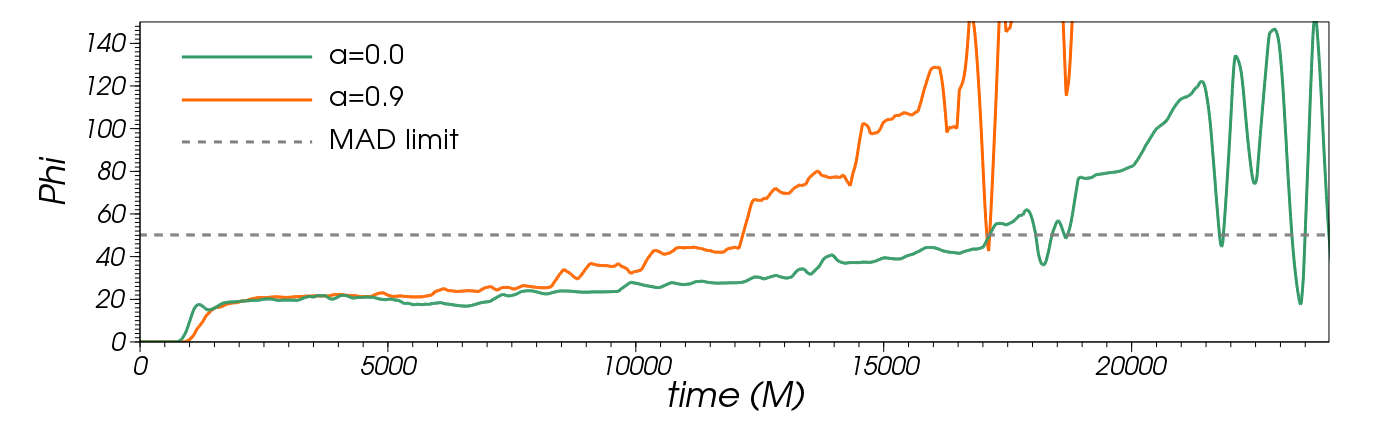}\\ \includegraphics[width=1.631\columnwidth]{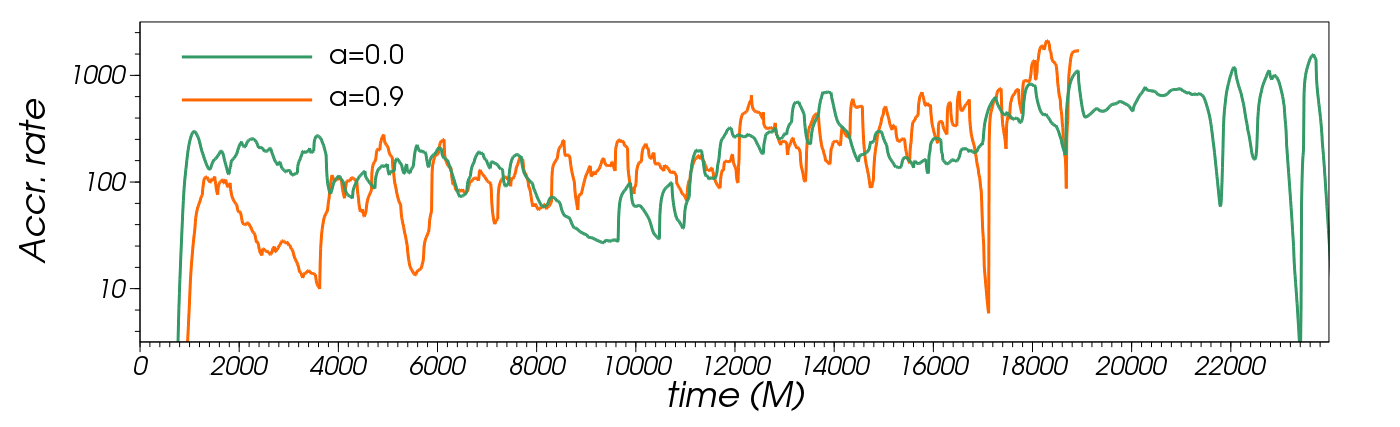}
\caption{Magnetic flux at BH horizon (top) and mass accretion rate in
  the Eddington units (bottom panel) for the $a_*=0.9$ model. The
  accretion rate was smoothen by calculating the moving average over $\Delta t=500$.}
  \label{f.r400phimdot}
\end{figure*}

Because of the large scale structure of the initial magnetic field in
the inner region of the torus, the magnetic flux on average keeps
accumulating at the BH.  The top panel of Fig.~\ref{f.r400phimdot} shows
the time evolution of the magnetic flux parameter $\phi$ defined as,
\be 
\phi = \frac 1{\sqrt{\langle\dot M\rangle}} \frac{4\pi}{2} \int_{0}^\pi
\int_0^{2\pi}\sqrt{-g}\,|B^r|\,{\rm d}\phi {\rm d}\theta, 
\ee 
where
$\langle\dot M\rangle$ represents the time-averaged accretion rate, $\dot M$, 
\be 
\dot M =
\int_{0}^\pi \int_0^{2\pi}\sqrt{-g}\,\rho u^r{\rm d}\phi {\rm
  d}\theta,
\ee 
averaged over time (see below), and $\sqrt{-g}$ is
the determinant of the metric.  The green and orange lines correspond
to $a_*=0.0$ and $a_*=0.9$, respectively.  The magnetic
flux at the BH is initially zero and starts to increase when gas
reaches the horizon. The initial increase around $t=1000$ is rapid
reflecting large magnetic pressure to gas density ratio near the inner
torus boundary. The magnetic flux increases much more slowly, however
steadily, later on. Around $t=17000$ (for $a_*=0.0$) and $t=12000$
($a_*=0.9)$ $\phi$ crosses the typical value for magnetically arrested
disks (MADs) $\phi_{\rm MAD}=50$ \citep{tchekh+12}. Such a state, when pressure of
the accumulated magnetic field balances the inward gravitational
force, cannot be maintained in two dimensions (in three dimensions gas
can distribute non-uniformly in azimuth and 'sneak in' between 
vertical magnetic field lines), and soon afterwards the accretion is
stopped and enters violent, unphysical evolution when gas bounces back
and forth from the magnetic wall at the horizon \citep{igu+03,mtb12}.  
This is clearly seen in
the bottom panel of Fig.~\ref{f.magtube2a} which shows the accretion
rate history.  Once the magnetic flux significantly exceeds the MAD
limit, the accretion rate violently oscillates. Because of this
limitation, we did not run the simulations longer and we limit the
subsequent analysis to the time period when the disk is far from the
MAD regime. We choose $t=2000\div15000$ and $t=2000\div12000$ for $a_*=0.0$ 
and $a_*=0.9$ runs, respectively. Such a relatively short duration of 
simulations is reflected in the fact that the extent
of the inflow equilibrium region is limited to, as discussed in the following
sections, to $R\lesssim 25$.

The accretion history plot shows the turbulent
character of the accretion and suggests average accretion rate
around $100\div200\dot M_{\rm Edd}$ for both runs. More detailed analysis is given below.

\subsection{Fluxes}

Looking at fluxes of rest mass and energy helps understand
the structure of an accretion flow. Fig.~\ref{f.400_2d}
presents (from top to bottom) time-averaged radial fluxes of rest 
mass ($\rho u^r$), 
radiation energy ($R^r_t$), magnetohydrodynamical energy ($T^r_t$), and energy available at infinity
($T^r_t+R^r_t+\rho u^r$). Left and right halves of the panels 
correspond to $a_*=0.0$ and $a_*=0.9$, respectively. Colors show
the magnitude of a given flux, arrows show its direction on
the poloidal plane. Solid white lines show the boundary between the inflow
(near the equatorial plane) and outflow (above) regions, which are
distinguished based on the sign of $\avg{\rho u^r}$\footnote{
We first average $\rho u^r$ and only then apply the outflow 
condition. \cite{yuan+12}, on the contrary, classified gas as outflowing
independently in each snapshot data. Discussion of these two methods 
is given in \cite{yuannarayan+14}.
}.
Dashed white lines denote the location of the photosphere ($\tau_{\rm tot}=1$).

\subsubsection{Rest mass flux}

\begin{figure}
\begin{tabular}{>{\centering\arraybackslash}m{0cm} >{\centering\arraybackslash}m{2cm}}
\begin{sideways}{\Large $\rho u^r$}\end{sideways}&
\includegraphics[width=1.01\columnwidth]{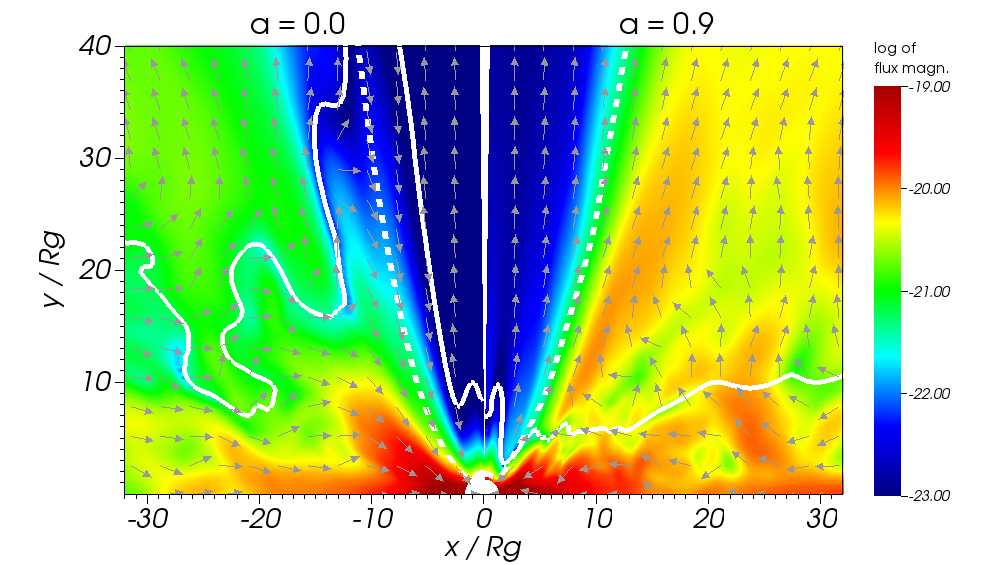}\\
\begin{sideways}{\Large $R^r_t$}\end{sideways}&
\includegraphics[width=1.01\columnwidth]{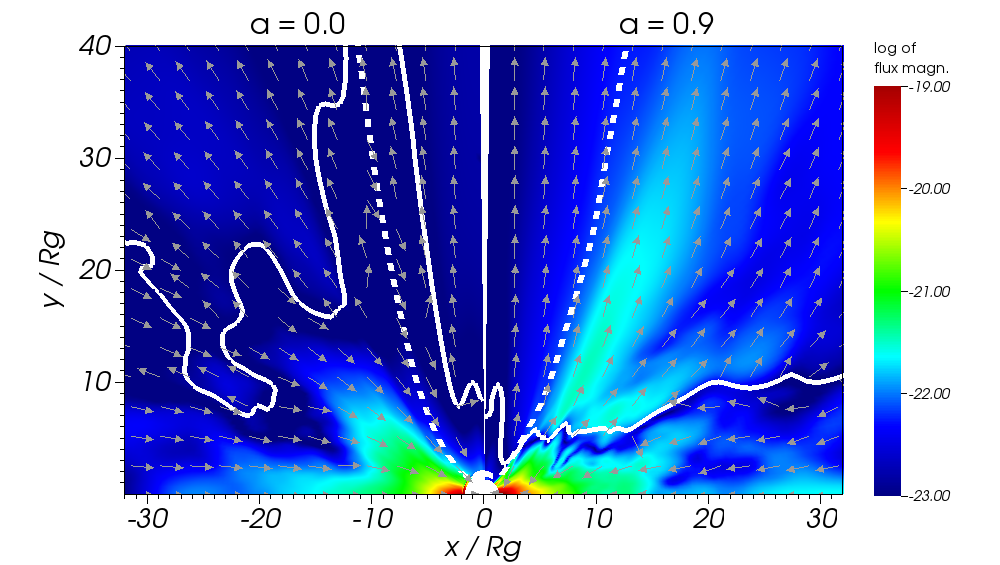}\\
\begin{sideways}{\Large $T^r_t$}\end{sideways}&
\includegraphics[width=1.01\columnwidth]{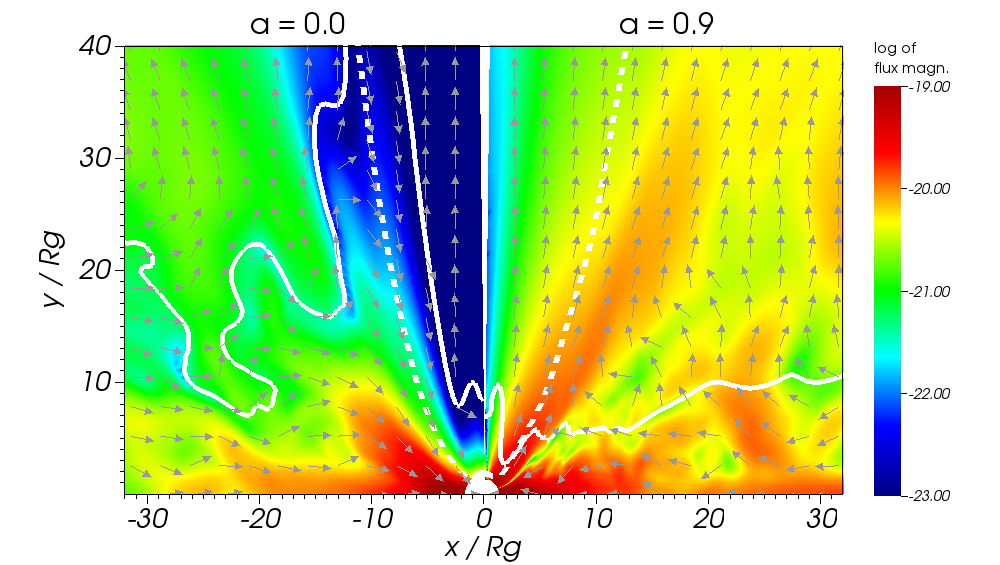}\\
\begin{sideways}{\Large $\rho u^r+T^r_t+R^r_t$}\end{sideways}&
\includegraphics[width=1.01\columnwidth]{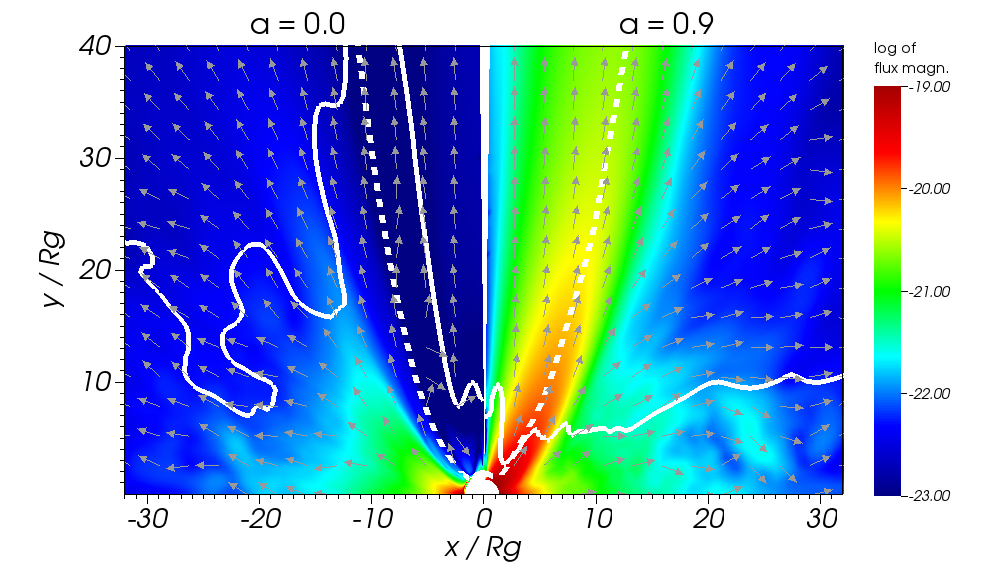}\\
\end{tabular}
\caption{Fluxes of rest mass ($\rho u^p$, topmost panel), total MHD energy 
($T^p_t$, second panel), radiative energy ($R^p_t$, third panel), and
the total energy minus rest mass ($T^p_t + R^p_t + \rho u^p$,
bottommost panel). Left and right halves correspond to $a_*=0.0$ and
$0.9$, respectively. Colors show the magnitudes of fluxes while arrows
show
their direction on the poloidal plane. The solid white line shows limits
the outflows region while the dashed one shows the photosphere ($\tau_{\rm tot}=1$).}
  \label{f.400_2d}
\end{figure}

The topmost panel in Fig.~\ref{f.400_2d} presents the flux of rest mass. Close to the
equatorial plane mass flows (on average) laminarly towards the BH.
In the case of a non-rotating BH the inflow region extends as far as
the photosphere. Mass flows out in the polar region, above the disk 
surface. However, the magnitude of the flux there is negligible.
The outflow region at intermediate polar angles at
radii $R>25$ is too far from the BH to reach the proper
inflow/outflow equilibrium and cannot be trusted.

The mass flow structure is qualitatively different for the
rotating BH case. Gas closest to the equatorial plane 
on average falls on 
the BH, as before. However, some fraction of the gas starts inside 
the inflow region, then crosses the solid white line and flows out. 
The region of outflows starts at the
polar axis and extends quite close to the equatorial plane
covering roughly $60\%$ of the total solid angle, $15\%$ of which 
at $R\approx 30$ is covered by the optically thin outflow in the polar funnel.
Outflowing gas accumulates roughly along polar angle $\theta=30^\circ$,
just outside the optically thin polar region, forming an optically thick 
wind.  This structure is specific only to rotating BH
and may be identified with the 'disk jet' found in 
numerical simulations of optically thin disks \citep{mtb12,tchekh+12,tchekh+12b,sadowski+outflows,yuannarayan+14}.

\begin{figure}
\hspace{-.0cm}\includegraphics[width=1.\columnwidth]{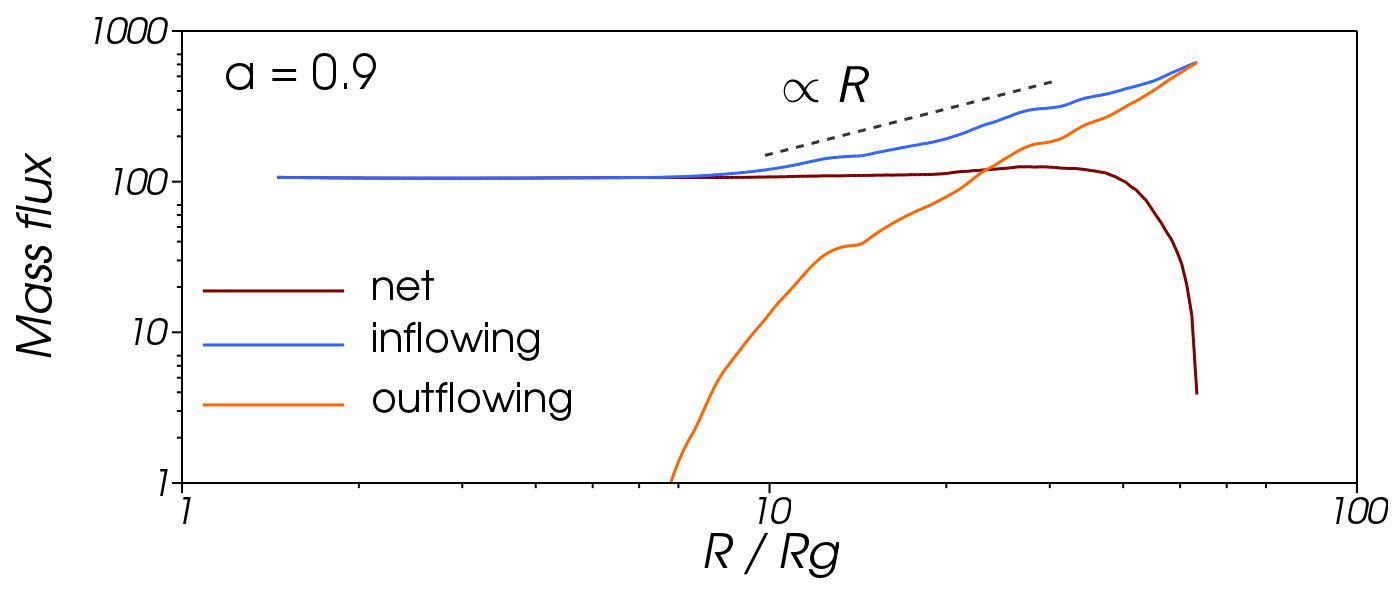}\\
\hspace{-.0cm}\includegraphics[width=1.\columnwidth]{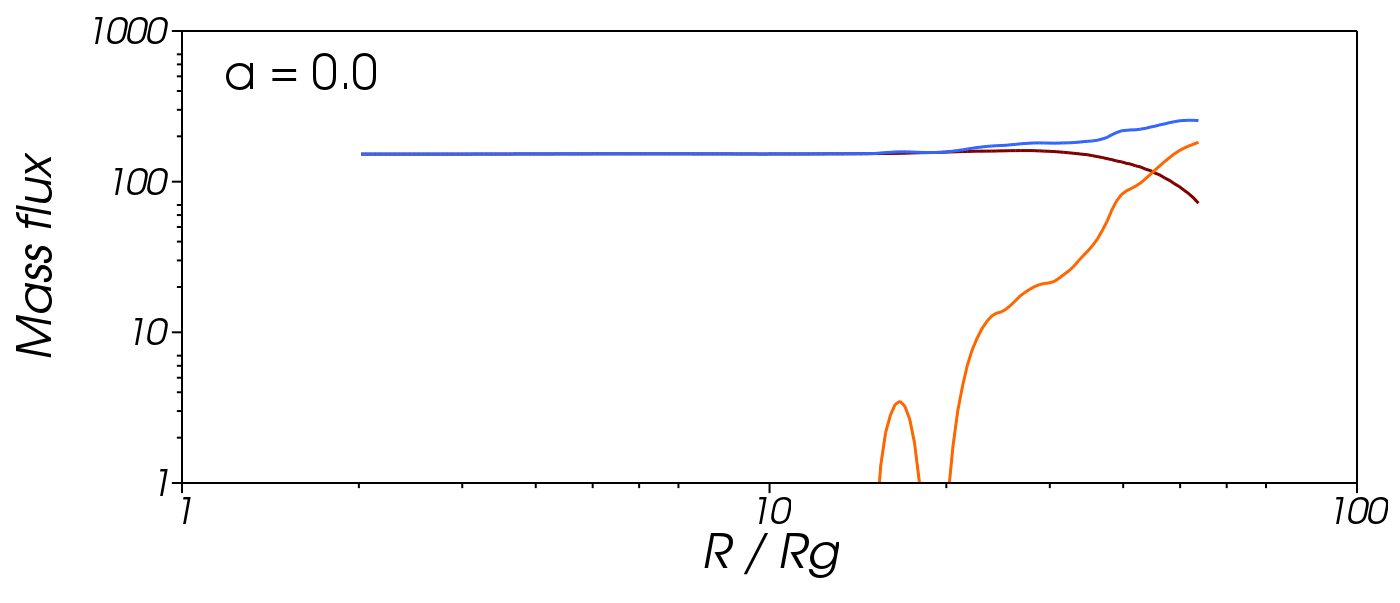}
\caption{Integrated rest mass flux in Eddington units (Eq.~\ref{e.mdotedd}) 
as a function of radius for
  $a_*=0.9$ (top) and $a_*=0.0$ (bottom panel) simulations. Three
  lines
are presented: net flux (red), mass outflow
($\avg{\rho u^r}>0$, blue) and
inflow rate ($\avg{\rho u^r}<0$, orange). }
  \label{f.400_mdotvsr}
\end{figure}

The magnitude of the outflow clearly is comparable with
the inflow accretion rate.
To quantify the mass flux we integrate
it over the total surface of a sphere and obtain the average net
flux, $\langle\dot M_{\rm net}\rangle$,
\be
\langle\dot M_{\rm net}\rangle=\int_{0}^\pi
\int_0^{2\pi}\sqrt{-g}\,\langle\rho u^r\rangle\,{\rm d}\phi {\rm d}\theta, 
\ee 
and correspondingly over the inflow and
outflow regions to obtain the integrated inflowing and outflowing
fluxes, respectively. Fig.~\ref{f.400_mdotvsr} shows the radial
profiles of integrated rest mass fluxes for $a_*=0.9$ (top) and
$a_*=0.0$ (bottom panel). The fluxes have been rescaled, and their
magnitudes are given in Eddington units ($\dot M_{\rm Edd}$).

The red lines in Fig.~\ref{f.400_mdotvsr} show the net flux. In steady
state it is expected to be constant in radius. In our simulations it
keeps a constant value with good accuracy up to $R\approx 25$ which
we consider the limit of inflow equilibrium in the bulk of
the disk. The average mass flux (accretion rate) is $\sim105$ and
$150\dot M_{\rm Edd}$ for $a_*=0.9$ and $a_*=0.0$, respectively.
The orange lines show the integrated flux of outflowing mass (the outflow
accretion rate). The outflows are significant only for the rotating 
BH -- in case of the BH with zero spin, the outflow emerges only around
and outside the radius of inflow equilibrium, and therefore is not
reliable. 
For $a_*=0.9$, the outflow starts around $R=10$ and
increases with radius following a power-law. Just outside radius $R=20$ the outflowing
flux equals the net flux, i.e., two parts flow in, one part flows out. 
The blue line shows the inflowing accretion rate, i.e., the sum of the two.
For our simulation with $a_*=0.9$, the inflowing accretion rate
decreases proportionally to the radius in the whole region
where outflows are not negligible ($R>10$). As a specific example,
the inflow accretion rate at $R=30$ is $\sim300\dot M_{\rm Edd}$.
Out of this only $\sim 100\dot M_{\rm Edd}$ reaches radius $R=10$, and
the remaining $\sim 200\Medd$ goes into an outflow. There is negligible
outflow inside $R\sim 10$, so the accretion rate on the BH is
$\sim 100\dot M_{\rm Edd}$.

\subsubsection{Radiative flux}

The second panel in Fig.~\ref{f.400_2d} shows the energy flux carried by photons, $R^r_t$. In the
optically thick region it qualitatively follows the distribution of mass
flux reflecting the fact that photons are trapped in gas when scattering
opacity is large. For this reason, there is no outflow of radiative energy inside 
the photosphere in $a_*=0.0$ simulation. The other run ($a_*=0.9$) shows
a region of outflowing radiative energy which exactly overlaps with the 
region of strong mass outflow (top panel). However, the amount of outflowing 
radiative energy is much smaller than the flux of energy flowing out in the
form of rest-mass in the same region. This is because gas temperature, which determines
the energy density of trapped photons through the LTE condition 
($\widehat E=4\sigma T^4$) is not relativistic.

\begin{figure}
%\hspace{-.0cm}\includegraphics[width=1.\columnwidth]{r400a9_Rdot.png}\\
%\hspace{-.0cm}\includegraphics[width=1.\columnwidth]{r400a0_Rdot.png}
\hspace{-.0cm}\includegraphics[width=1.\columnwidth]{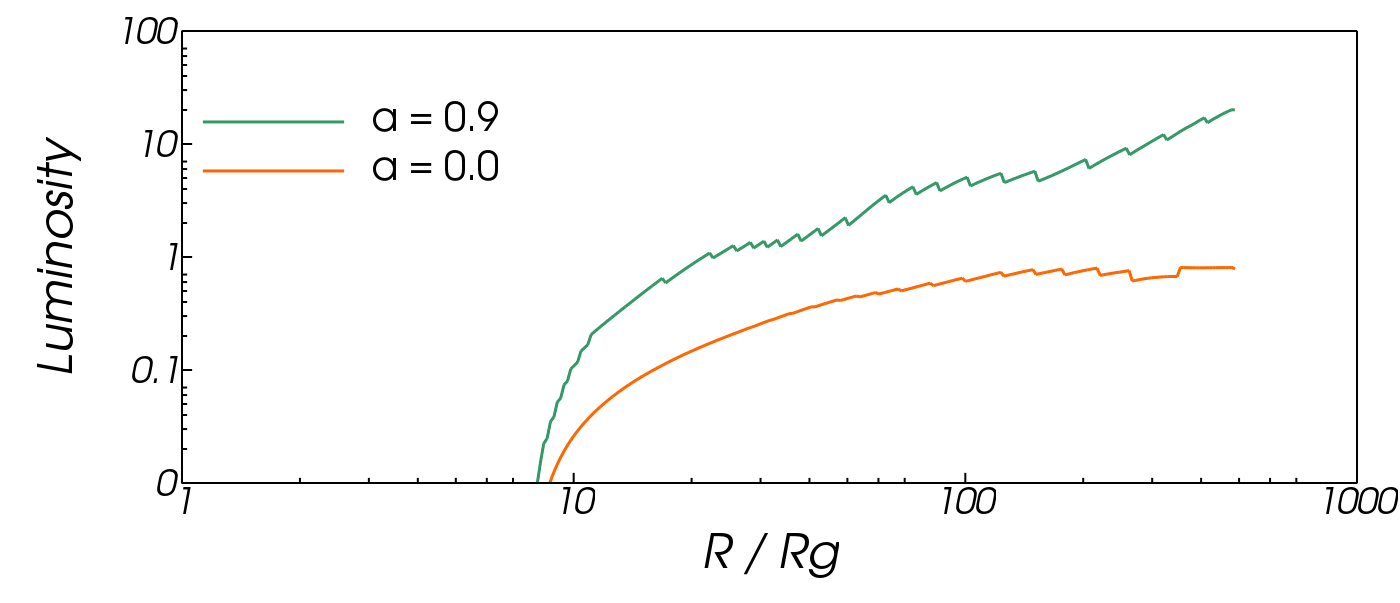}
\caption{Radiative flux of energy integrated over the
optically thin funnel in the Eddington units (Eq.~\ref{e.ledd}) as a function of radius for
  $a_*=0.9$ (green) and $a_*=0.0$ (orange line).}
  \label{f.400_rdotvsr}
\end{figure}

Fig.~\ref{f.400_rdotvsr} presents 
radiative flux of energy integrated over the optically thin
funnel, and normalized to the Eddington
luminosity,
\be
\label{e.ledd}
L_{\rm Edd}=1.25\times 10^{38} \frac{M}{M_\odot} \,\rm erg/s. 
\ee
The green and orange lines correspond to BH spins
$a_*=0.9$ and $0.0$, respectively.
The luminosity leaving
the disk along the polar axis in the optically thin funnel
increases with radius reflecting the fact that 
radiation diffuses into this region from surface layers of the optically
thick region. The gas that releases this radiation is located just inside the photosphere,
and, for the $a_*=0.9$ disk,
flows out from the equilibrium region nearest to the BH with relatively high velocity $\sim 0.1c$.
Therefore, the radiation field in the funnel is in the 
equilibrium. Near the outer boundary the luminosity
carried away by photons in this region is 
$\sim 10.0L_{\rm Edd}$ and $\sim 0.7L_{\rm Edd}$ for 
$a_*=0.9$ and $a_*=0.0$, respectively.

However, most of the radiative flux for $a_*=0.9$
flows out trapped in the gas in the optically thick region (see the second panel in Fig.~\ref{f.400_2d}).
The total radiative luminosity of such trapped photon flux in the outflow region
is $\sim 70L_{\rm Edd}$ already at $R=20$ and also increases with radius as a result
of gas outflowing from larger radii. It is not clear
what fraction of this luminosity can reach the observer.
Radiative energy trapped in the gas may contribute
to its thermal and kinetic energy before the gas expands
enough to become optically thin and release the photons \citep{poutanen+07}.
Therefore, the luminosity in the optically thin region
is only the lower limit for the total radiative luminosity of the disk.

\subsubsection{Energy flux}

The third panel from the top in Fig.~\ref{f.400_2d} shows the flux of 
MHD energy $T^r_t$ which includes 
the rest mass, gravitational, kinetic, and magnetic energies. For
the non-rotating BH (left half) no other form of energy is comparable 
with the rest-mass energy and the distribution of $T^r_t$ closely
resembles that of $\rho u^r$. The same is true for the rotating BH
but only far from the polar axis. In the polar region, on the contrary,
the flux of magnetic energy exceeds significantly the 
rest mass flux. It results from the Poynting flux which fills the polar (jet) region
presumably as a consequence of the \cite{blandfordznajek-77} process.

Energy flows out as 
rest-mass, MHD and radiative energy. However, only MHD and radiative 
energies represent usable energy at infinity. Therefore,
the quantity of interest from the point of view of the efficiency of
BH feedback is the total energy minus rest-mass energy. We call this
quantity ``energy available at infinity'', and define its flux as
 $\dot e_\infty=T^r_t+R^r_t+\rho u^r$ (the plus sign in front of $\rho u^r$ 
reflects negative signs of the other two components). The bottommost
panel in Fig~\ref{f.400_2d} shows the magnitude of this flux averaged
over time. 

The left half shows results from the non-rotating BH run. The energy is extracted 
inside the  optically thick bulk of the disk. This positive (towards infinity)
energy flux results from three factors. Firstly and most importantly,
gas looses its angular momentum on the way towards the BH and transfers it 
outward effectively transporting kinetic energy out.
Secondly, gas falling in strong gravitational potential heats up to
conserve the total energy. If disk was optically thin, this extra internal
energy would fall with the gas on the BH. For optically thick disks, however,
gas can cool and the extra energy can be put in the radiation field and
diffuse out. Thirdly, the accreting gas was initially
slightly bound --- the initial torus had a typical relativistic 
Bernoulli parameter \citep{narayan12} $Be\approx-0.02$, and accretion of a bound
(negative energy) gas effectively deposits positive energy at infinity. 
The initial value of the Bernoulli parameter determines
the uncertainity of the energy flux estimates.

The structure of the flux of energy available at infinity for the rotating BH
is qualitatively different. It is dominated by the flux of magnetic energy
flowing out along the polar axis - the jet. The amount of energy transferred 
in this way exceeds significantly all the contributions discussed above which
do not rely on BH rotation. The outflow originates at the BH and apparently
extracts
its rotational energy through
the Blandford-Znajek mechanism. 

\begin{figure}
\hspace{-.0cm}\includegraphics[width=1.\columnwidth]{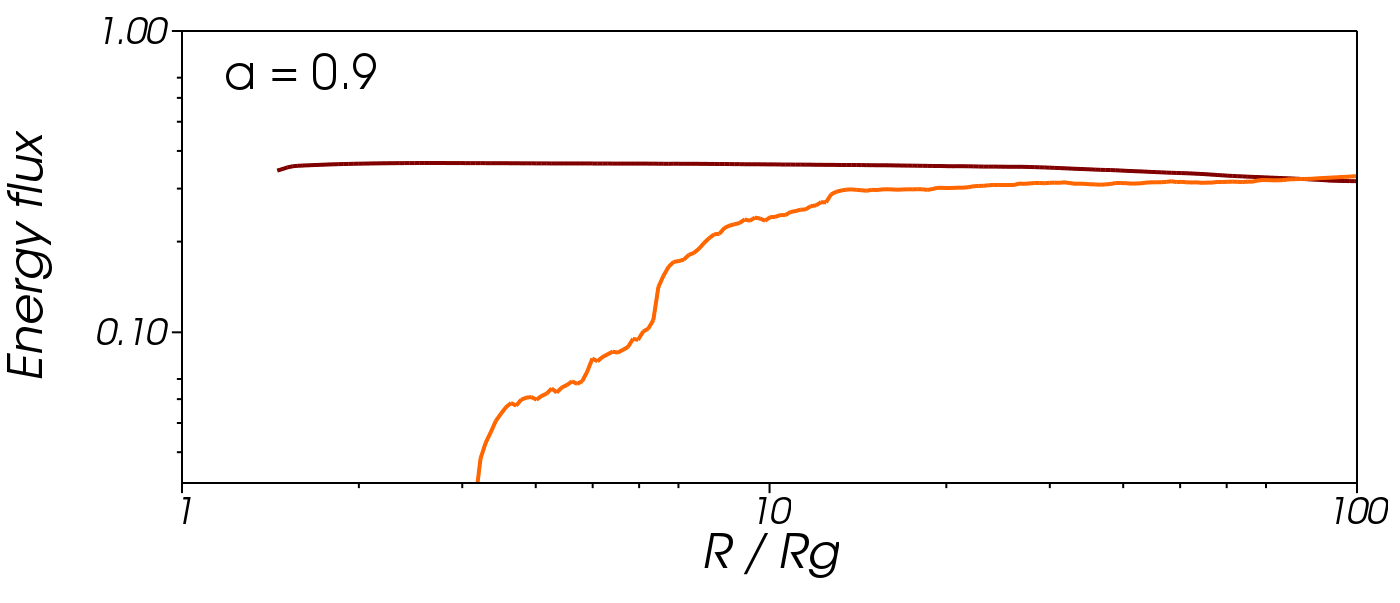}\\
\hspace{-.0cm}\includegraphics[width=1.\columnwidth]{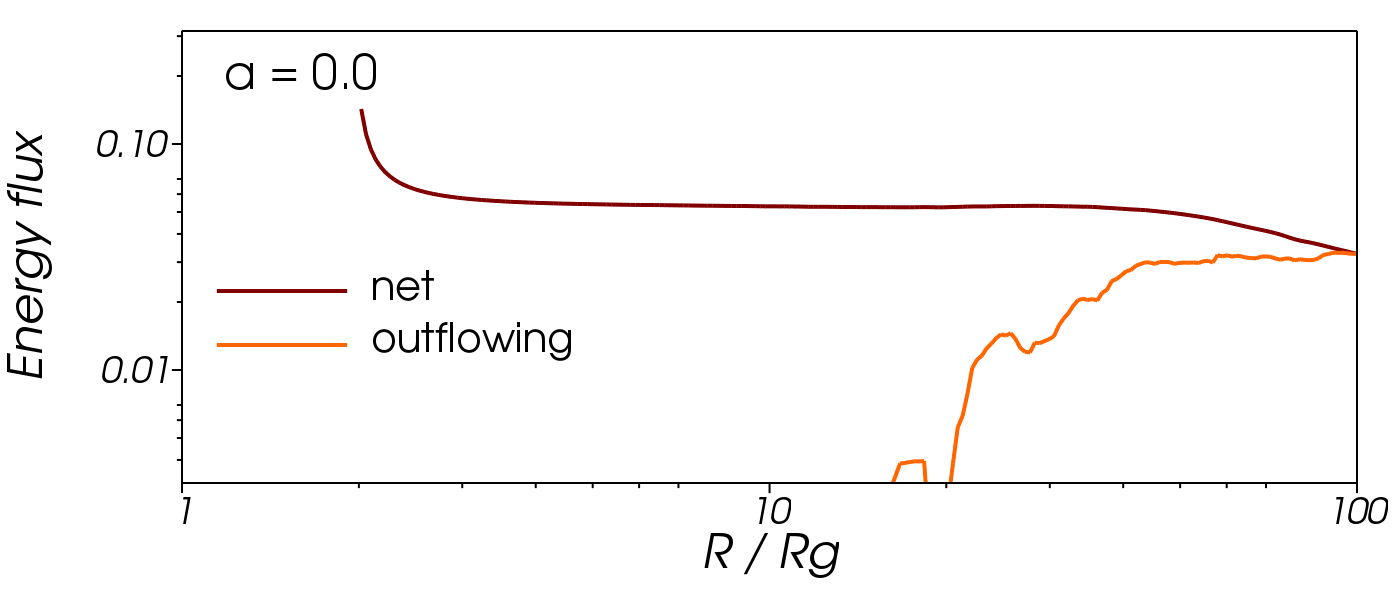}
\caption{Similar to Fig.~\ref{f.400_mdotvsr} but for the total energy
  flux,
$\avg{T^r_t}+\avg{R^r_t}+\avg{\rho u^r}$, normalized to the rest mass flux $\langle\dot M_{\rm net} c^2\rangle$ at the BH.}
  \label{f.400_edovsr}
\end{figure}

In Fig.~\ref{f.400_edovsr} we plot fluxes of energy at infinity
integrated over spherical surfaces and normalized by the average 
net rest-mass energy flux ($\langle\dot M_{\rm net}\rangle$), e.g.,
\be
\label{e.Edotnet}
\langle\dot E_{\rm net}\rangle=\frac{1}{\langle\dot M_{\rm net}\rangle}\int_{0}^\pi
\int_0^{2\pi}\sqrt{-g}(\langle T^r_t\rangle+\langle R^r_t\rangle+\langle \rho u^r\rangle)\,{\rm d}\phi {\rm d}\theta.
\ee 
Red and orange lines correspond
to fluxes integrated over the whole surface (net flux) and over the
region of outflowing gas, respectively. The net flux is reasonably constant
up to $R\approx 30$ indicating the inflow/outflow equilibrium in this region.
The departure from constant value near the horizon, most significant for the $a_*=0.0$
run, reflects the impact of artificial floors in density imposed there.
The amount of energy available at infinity that flows out of the system 
is roughly $0.33\dot M c^2$ and $0.05\dot M c^2$ for $a_*=0.9$ and $a_*=0.0$ runs,
respectively. The orange line, denoting the corresponding flux in the region
of outflowing gas, is close to the total flux in the outflow region
what reflects the fact that the energy carried in by the inflow is
small when compared to the energy of the outflow.

Previous works \citep{tchekh10b,tchekh10a,tchekh11,tchekh+12,narayan+10,penna+13b,sadowski+outflows} have analyzed the power of jets produced by
rotating BHs in the presence of geometrically thick, optically thin disks and showed that it
was consistent with the Blandford-Znajek
process which predicts the jet power proportional to the square of
magnetic flux at BH horizon, $\phi$, and increasing with BH spin.
The average value of $\phi$ for the $a_*=0.9$ simulation within
the time window we analyze, i.e., between $t=2000\div12000$ is
$\avg\phi\approx 25$ (Fig.~\ref{f.r400phimdot}). For such values
of the magnetic flux and BH spin the empirical formulae
from \cite{sadowski+outflows} predict the total (jet plus wind) 
amount of energy available at infinity to be $0.32\dot Mc^2$,
which is close to the value $0.33\dot Mc^2$ we get in the current simulation.
This fact indirectly proves that the jet power comes from the
BH rotational energy extracted through the Blandford-Znajek
process, and that the presence of radiation hardly modifies the 
jet properties when compared to optically thin disks.

\subsection{1D profiles}

\begin{figure}
\hspace{-.0cm}\includegraphics[width=.9\columnwidth]{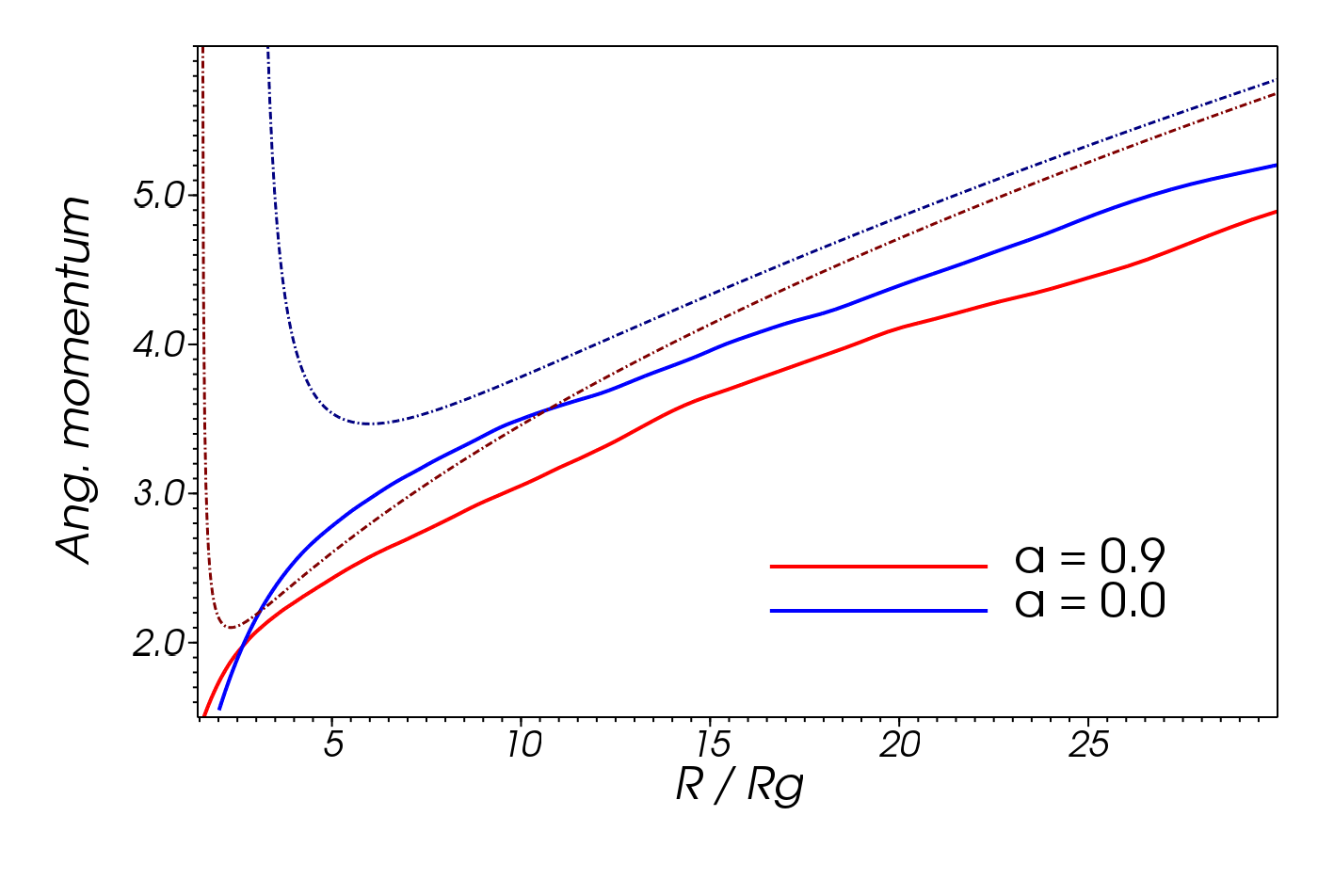}\\
\caption{Angular momentum of gas, $\langle u_\phi\rangle$ (Eq.~\ref{e.uphi})
for $a_*=0.0$ (blue), and $a_*=0.9$ (red line) simulations. The dashed 
lines show corresponding profiles of Keplerian angular momentum. }
  \label{f.uphi}
\end{figure}

In this Section we discuss vertically averaged physical quantities
characterizing the flow. We start with the density-weighted angular momentum of gas
(Fig.~\ref{f.uphi}), 
\be 
\label{e.uphi}
\langle
u_\phi\rangle=\frac1{\langle\Sigma\rangle}\int_0^{\pi/2} \langle\rho
u_\phi\rangle \sqrt{g_{\theta\theta}}\,{\rm d}\theta, \ee 
where
$\langle\Sigma\rangle=\int_0^{\pi/2}\langle\rho\rangle\sqrt{g_{\theta\theta}}\,{\rm
  d}\theta$ stands for the surface density. The blue and red lines
correspond to non-rotating and spinning BHs. The dashed lines 
show corresponding Keplerian angular momentum profiles.
For both simulations the angular momentum is sub-Keplerian. In 
the region outside the ISCO, the angular momentum is on average $15\%$
below the Keplerian value. Close to the BH the angular momentum
rapidly drops. It does not mean, however, that there is 
torque exerted on the gas at the horizon because $u_\phi$ is 
not a constant of motion for a magnetized fluid, especially, in 
the presence of radiation field.

\begin{figure}
\hspace{-.0cm}\includegraphics[width=.9\columnwidth]{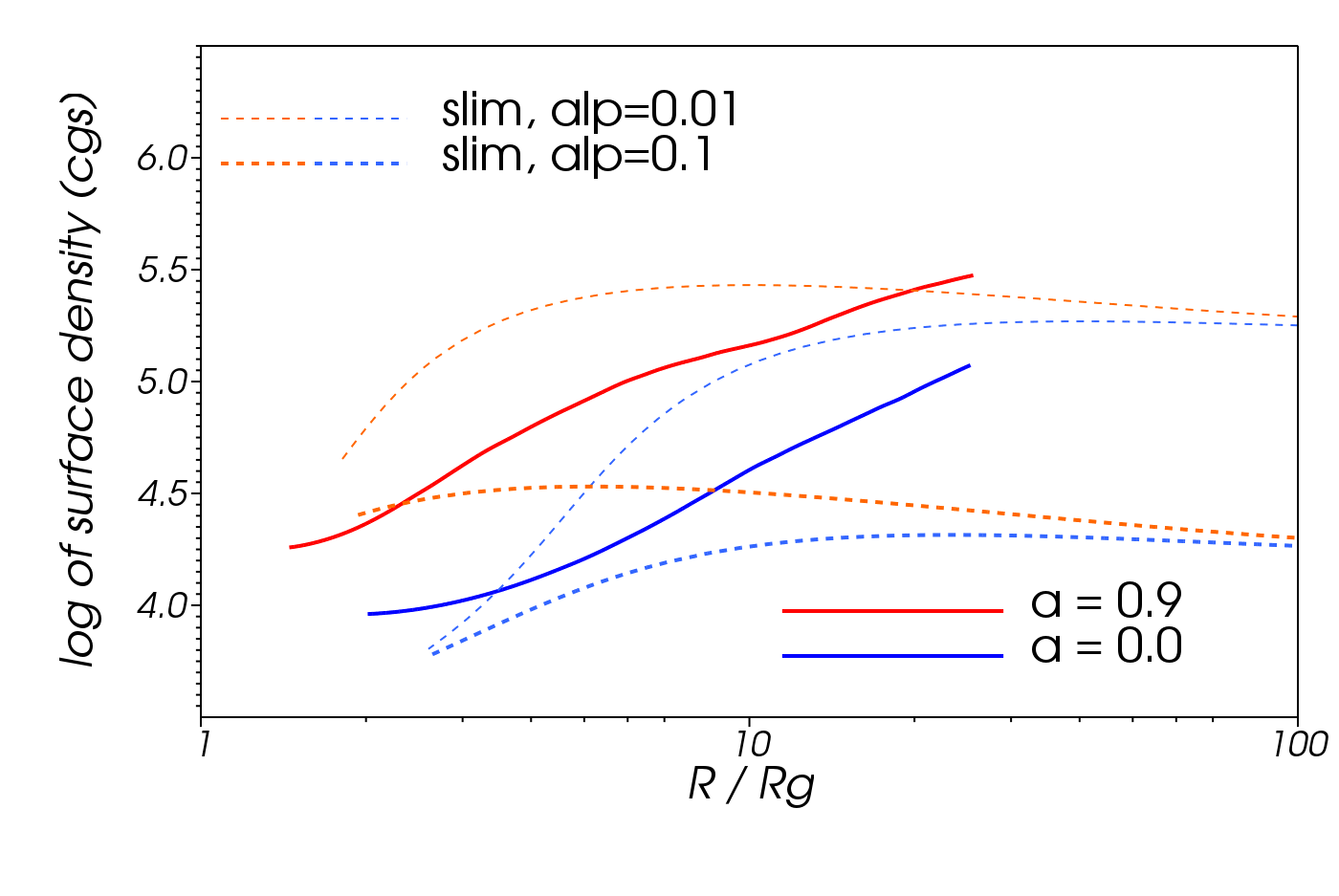}\\
\hspace{-.0cm}\includegraphics[width=.9\columnwidth]{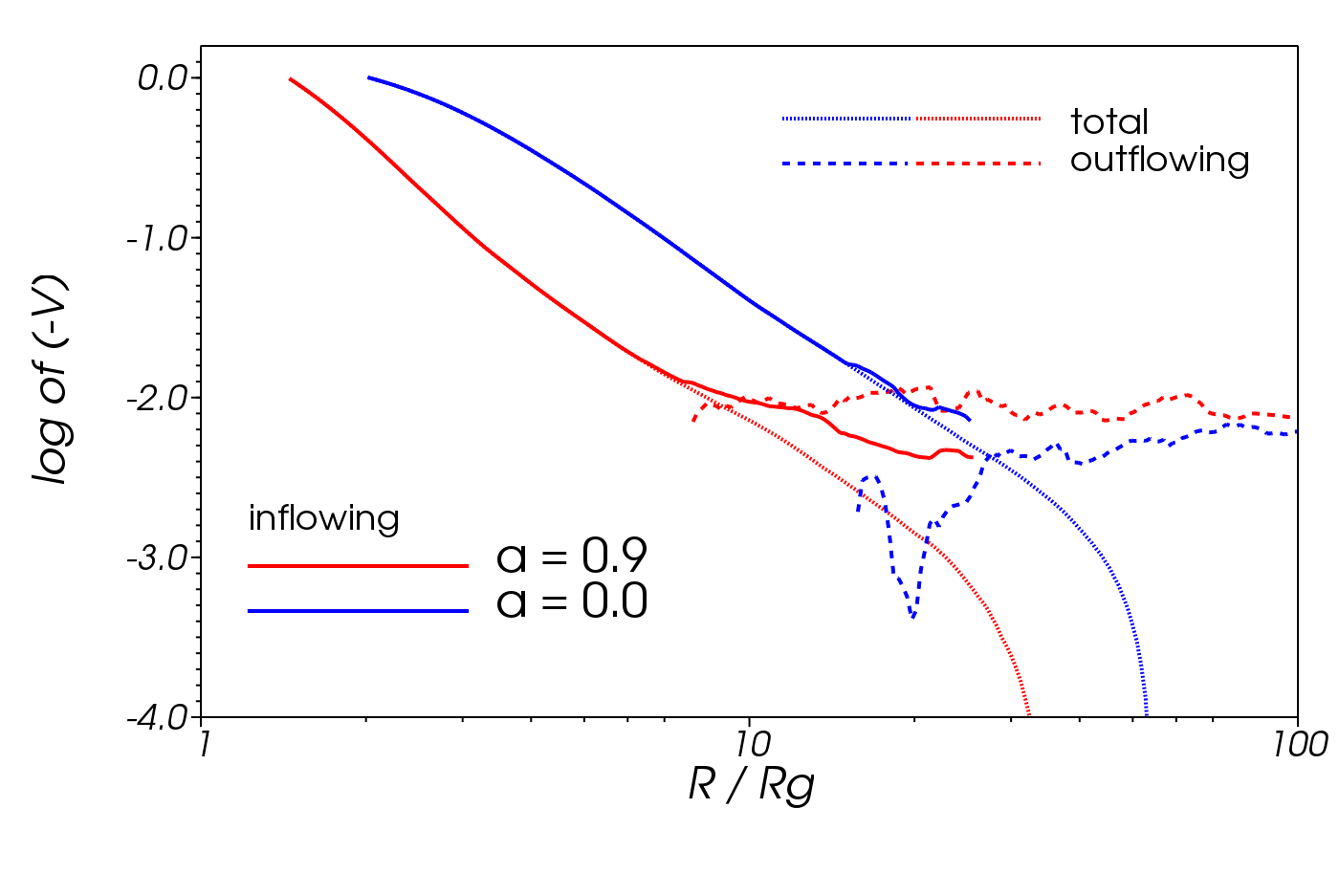}
\caption{Top: Surface density (in $g/cm^3$) of the simulated disks (thick lines) and
  of slim disks \citep{sadowski.phd} with $\alpha=0.01$ (thin) and 
$\alpha=0.1$ (thick dotted lines). Bottom: Radial velocity profiles
for the inflowing gas (thick), all the gas (thin), and the outflowing
gas (dotted lines).}
  \label{f.sigmavrad}
\end{figure}

Fig.~\ref{f.sigmavrad} shows the average surface density, $\langle \Sigma\rangle$ (top), 
and the average radial velocity as measured by an observer corotating with the 
fluid \citep{abramowiczetal95}, $V$ (bottom panel), given by
\be
\avg{V}^2 = \frac{\avg{u^r}\avg{u_r}}{1+\avg{u^r}\avg{u_r}},
\ee
where $\avg{u^r}$ is the density-weighted radial component of contravariant velocity,
\be
\label{e.ur}
\langle
u^r\rangle=\frac1{\langle\Sigma\rangle}\int_0^{\pi/2} \langle\rho
u^r\rangle \sqrt{g_{\theta\theta}}\,{\rm d}\theta.
\ee 
We choose to show $\avg{V}$ because it equals $1$ at the horizon. 

In the top panel we show surface density from the simulation data with
thick lines and terminate them
at radius $R=25$ corresponding to the extent of the inflow equilibrium
in the disk. In addition,  we plot surface 
density profiles from the relativistic slim disk model by \cite{sadowski.phd} for
$\dot M=100\dot M_{\rm Edd}$ and two values of the viscosity parameter $\alpha$: 
$0.01$ (thin) and $0.1$ (thick dashed lines). The surface density in our simulations drops
monotonically towards the BH reaching $\sim 10^4\,\rm g/cm^3$ near the horizon. The
slim disk profiles predict the same order of magnitude of surface density, however 
its shape is different from the numerical result. This reflects the
fact that the slim disk model does not account for the magnetic
pressure which is significant in the innermost region and which 
modifies disk structure and dynamics there. 

The radial velocities of the simulated disks are shown in the bottom panel.
This time the thick solid line corresponds to the radial velocity $V$ averaged
over the inflow region (the disk bulk). The dashed and  dotted thin lines show velocity
profiles for the outflow region and the whole simulation, respectively.
The latter roughly follows a single
power-law, $\avg{V}\propto R^{-2}$ up to certain radii, $R=30$ and $50$ for
$a_*=0.9$ and $0.0$, respectively, where the velocity averaged over the whole 
domain becomes positive
due to prevailing outflows emerging from inside this radius --- these
radii correspond to intersections of the 'net' and 'outflowing'
curves in Fig.~\ref{f.400_mdotvsr}. The solid lines correspond to the
velocity averaged over the inflow region (the disk bulk). Its profile
flattens in the region of non-zero outflows (e.g., $R>10$ for $a_*=0.9$) and 
seems to approach $R^{-1}$ slope once outflows are significant. The dashed lines 
show the average (positive) radial velocity in the outflow region. The
outflowing gas is accelerated to relatively uniform radial velocity $V\approx 0.01$,
similar in order of magnitude to the midly-relativistic 
winds observed in some AGN \citep{tombesi+10a,tombesi+10b,weng+13}.

\section{Summary and Discussion}
\label{s.summary}

In this paper we have presented a magnetohydrodynamic (MHD)
generalization of the GR radiation hydrodynamic code \koral\, described
in a previous paper \citep{sadowski+koral}. The code evolves the ideal
MHD equations of magnetized gas in a fixed relativistic metric,
evolving radiation in parallel as a second fluid that is coupled to
the gas via scattering and absorption processes.  The radiation fluid
is described by the radiative energy density and radiative flux, from
which M1 closure enables us to calculate (approximately) all the
second moments of the radiation intensity. We have carried out a
number of tests and verified that the code correctly treats the
dynamics of magnetized gas and handles the coupling between gas and
radiation in both the optically thin and optically thick limits.

The M1 closure scheme employed here is approximate in the sense that
\koral\, does not include any frequency dependence, but instead evolves
the frequency integrated radiative energy density and flux, with grey
opacities. Nevertheless, this work constitutes a significant
improvement over previous work since the version of \koral\,
described here is capable of evolving magnetized gas
and radiation in a GR space-time. In addition, the M1 closure scheme
is superior to both Eddington closure and flux-limited diffusion, two
approaches generally favored in many previous (non-GR) radiation codes
\citep[e.g.,][]{ohsuga09,ohsuga11,zanottietal11}, 
though it is not as sophisticated as the (non-GR)
variable Eddington tensor approach described in \cite{jiang+12}.

While the primary purpose of this paper is to describe the code and
establish its performance, in \S5 we present an initial application in
the area of black hole accretion.  We have performed two axisymmetric
simulations of super-critical (i.e., super-Eddington: $\dot{M} >
\dot{M}_{\rm Edd}$) accretion, one corresponding to a non-rotating BH
($a_*=0$) and one to a rotating BH ($a_*=0.9$). The accretion rates in
the two simulations are highly super-Eddington, $105 \dot M_{\rm Edd}$
and $150 \dot M_{\rm Edd}$, respectively.  Both simulations are
initialized with a torus of radiation-dominated orbiting gas with a
weak poloidal magnetic field. The MRI develops promptly and gas
settles quickly into a quasi-steady, geometrically and optically
thick, sub-Keplerian, accretion disk. Twin funnels form spontaneously
along the polar axis and serve as channels through which gas, magnetic
field and radiation escape relativistically.  Surrounding the funnel
is a more wide-angle, less relativistic, outflow.

Our results for a non-rotating BH are in qualitative agreement with
models A in \cite{ohsuga09} and \cite{ohsuga11}, which were simulated
using a non-relativistic MHD code with flux-limited
diffusion. Detailed comparison is difficult because the accretion rate
in those models is $\sim6\Medd$ (in our units) compared to $105\Medd$
in the model described here. In the case of our $a_*=0.9$ model, there
is no previous comparable simulation. Relativity introduces
qualitatively new effects for a spinning BH --- frame-dragging,
Penrose process, BZ effect --- which only a GR code can properly
explore. Many GRMHD simulations without radiation have been described
in the literature 
\citep[e.g.,][]{devilliersetal03,gammie03,anninosetal05,delzannaetal07,
shafee08,penna10,
tchekh10b,tchekh10a,narayan+10,nobleetal11,tchekh11,tchekh+12,
narayan+12b,sadowski+outflows}, and we compare our results to some of that
work below.  However, there is no previous GRMHD simulation with
radiation.

The first, perhaps trivial, result from the simulations reported here
is that super-Eddington accretion on to a BH is permitted, and will
occur whenever there is adequate mass supply on the outside. The two
simulations in this paper were initialized with a large amount of mass
%(\AS{can we estimate the torus mass in units of $Mdot_Edd$ times
%  viscous time at torus pressure max?}) 
in a torus relatively close to
the BH with inner edge at $R=15$.  This initial state is probably not
dissimilar to what is expected in tidal disruption events,
where many models predict super-Eddington accretion with $\dot{M}$ as
high as $100\dot{M}_{\rm Edd}$ early in the outburst \citep{krolikpiran-12,tchekh+swiftmad}.

We
find considerable mass loss
in the $a_*=0.9$ run, as expected for a geometrically thick accretion disks \citep[e.g.,][]
{narayanyi-94,narayanyi-95,blandfordbegelman-99,stone+99,
stonepringle-01,yuan-01,narayan12}. On the contrary, the
$a_*=0$ simulation has no significant outflow within the inflow
equilibrium radius of $R\approx 25$. Crudely, it appears that the
mass inflow rate in the $a_*=0.9$ simulation scales as $\dot{M}_{\rm in} \sim
\dot{M}_{\rm BH} (R/10)^s$ for $R\geq10$ and $\dot{M}_{\rm in} =
\dot{M}_{\rm BH}$ for $R<10$, with $s\sim1$, though we are not
confident of the exact value of this index. In any case, for mass
injection at a radius of a few tens of gravitational radii, mass loss
introduces at most a factor of a few difference between the amount of
mass that is supplied and the amount that is accreted. Thus, for tidal
disruption events, to a good approximation, most of the gas does end
up on the BH.

For accretion from a much larger distance, e.g., a supermassive BH
accreting at a highly super-Eddington rate from a surrounding
interstellar medium \citep[e.g.,][]{siemiginowska+10,netzer+13}, or a stellar-mass BH like SS433
hyper-accreting from a companion star \citep{fabrika-04}, there could be
considerable mass loss via an outflow. From the above scaling, the
mass that accretes on the BH is only $\sim (10/R_{\rm outer})^s
\dot{M}_{\rm outer}$, where $\dot{M}_{\rm outer}$ is the mass supply
rate at the outer radius $R_{\rm outer}$. Nevertheless, provided there
is a sufficiently large mass supply at the outer radius, which is
perhaps not so unlikely early in the life of a gas-rich proto-galaxy
at high redshift, super-Eddington accretion should readily occur.

The simulations show that super-critical accretion is radiatively
inefficient. Although the accretion rate exceeds the Eddington rate by
two orders of magnitude, the photospheric luminosity is only of order
a few $L_{\rm Edd}$.  At a radial distance of $R\sim100$ within the
funnel, where we believe the results are fairly trustworthy, the
radiative luminosity of the $a_*=0$ model is $\sim0.6L_{\rm Edd}$ and
that of the $a_*=0.9$ model is $\sim5L_{\rm Edd}$
(Fig.~\ref{f.400_rdotvsr}). Some additional radiation is likely to be
emitted at larger radii in the funnel, and we also expect of order
$1L_{\rm Edd}$ of luminosity to escape from the outer photosphere of
the accretion disk itself.  But all in all, it appears that the total
luminosity is limited to a $\sim {\rm few}-10 L_{\rm Edd}$. Writing
the radiative luminosity, $\eta$, as $L = \eta \dot{M}_{\rm BH}c^2$, this
corresponds to $\eta\approx 10^{-3}$, i.e., the flow is
radiatively very inefficient. Since the simulations correspond to an
advection-dominated regime of accretion, this result is expected.

Nevertheless, one interesting result is the unusually large radiative
luminosity emitted through the narrow funnel. Out at $R=100$, the
funnel opening angle (measured out to the photosphere) is only
$\sim0.2$\,rad, and at $R=500$ it is even smaller $\sim0.1$\,rad.  Yet
the luminosity emitted through the funnels is greater than Eddington
and perhaps as much as $10L_{\rm Edd}$. Expressed as a radiative flux,
the simulated models are thus enormously super-Eddington in the
funnels: $F_{\rm funnel} \sim 10^3-10^4 F_{\rm Edd}$. 
This result confirms ideas developed decades ago
regarding super-Eddington luminosities in funnels \citep[e.g.,][]{sikora-81,
narayan+83}. Relativistic beaming and
relativistic gas motions are at least part of the explanation for
these large radiative fluxes. In the context of the tidal disruption
event Swift J1644+57 \citep{bloom+11,levan+11,zauderer+11,berger+12}, models usually require very large ``isotropic
equivalent luminosities'' at early times. The present simulations
indicate that, for an observer who is favorably placed with respect
to the funnel, such luminosities are easily achieved.

The luminosity estimate given above refers only to the radiation in
the topically thin region of the funnel. This is the radiation that we
can be sure will escape. There is in addition a substantial volume of
the optically thick outflow where there is a net outward flux
of radiative energy. The corresponding luminosity is about a factor of
10 larger. However, this radiation is trapped in the gas and
participates in the overall outflow and expansion of escaping gas. How
much of this radiation will finally flow out of the photosphere and
how much will be used to do work on the expanding gas is not easy to
estimate. Hence we do not include this component of the radiation flux
in our luminosity estimates.

We can, however, estimate the {\it total} energy loss rate from the
system, viz., the quantity $\avg{\dot{E}_{\rm net}}$ (Eq.~\ref{e.Edotnet}), which measures the
combined energy flux to infinity in the form of radiation, magnetic
field, and gas thermal and kinetic energy. This energy loss includes
energy outflow in both the optically thin and optically thick regions,
and is quite substantial.  In fact, we see no sign of any
``inefficiency'' in the overall energy loss rate from the system. The
$a_*=0$ simulation has an energy loss rate equal to 5\% of
$\dot{M}_{\rm BH}c^2$, while the $a_*=0.9$ simulation loses energy at
33\% of $\dot{M}_{\rm BH}c^2$. Surprisingly, these energy outflow
efficiencies are quite similar to those estimated for hot
non-radiative accretion flows \citep{tchekh11,sadowski+outflows}. 
It appears that scalings derived for those systems may be
applied also to super-Eddington accretion flows. This result may have
large implications for BH feedback in early stages of galaxy
formation in the universe.

If the above result is confirmed, it would imply that the jet
efficiency is insensitive to whether accretion occurs via a
geometrically thick non-radiative hot accretion flow or a
geometrically thick radiation-trapped super-Eddington flow. What seems
to matter is the BH spin and the magnetic flux around the BH, exactly
as expected with the \cite{blandfordznajek-77} model. The demonstration
in this paper of a BZ-like mechanism is likely operating in the
radiation-dominated accretion regime is a new result. In the case of
hot accretion flow simulations, recent work has shown that the
simulated jets follow the BZ model very well \citep{penna+13b} and
that the simulations can be understood as a generalized Penrose
process \citep{lasota+13}. This explains how it is possible for jet
efficiencies to exceed 100\% in some cases \citep{tchekh11}. 
A similar demonstration in the case of radiation-dominated
accretion flows would be very worthwhile.

In the context of hot accretion flows, the above work of
\cite{tchekh11} has revived interest in the magnetically arrested disk
(MAD) model, in which a very strong magnetic field accumulates around
the BH and disrupts the accretion flow. MAD systems produce extremely
powerful jets because of the very strong magnetic field around the BH.
Both the simulations described in the present paper reach the MAD
limit -- at a time of $17000M$ for $a_*=0$ and $12000M$ for $a_*=0.9$.
Beyond this point, these axisymmetric simulations cannot be trusted
since the MAD regime corresponds to intrinsically non-axisymmetric
flows. Future 3D radiation GRMHD simulations can explore this regime.
For now, what can be said is that the MAD regime appears to be as easy
to achieve in super-Eddington flows as in low-$\dot{M}$ hot accretion
flows. This again has implications for jet power and AGN feedback.

\section{Acknowledgements}

We thank Juri Poutanen for useful discussions.
RN and AS were supported in part by NASA grant
NNX11AE16G.  We acknowledge NSF support via XSEDE
resources under grant numbers TG-AST080026N
(RN and AS) and TG-AST100040 (AT), and NASA support via High-End
Computing (HEC) Program through the NASA Advanced Supercomputing (NAS)
Division at Ames Research Center (AS and RN).
AT was supported by a Princeton Center for Theoretical Science
Fellowship 
and by NASA through the Einstein Fellowship Program, grant PF3-140115.
 
\bibliographystyle{mn2e}

\appendix

\section{Radiative shock near the polar axis}
\label{ap.a}

At the onset of the simulations, the radiative flux supporting
the torus against the effective gravity starts to escape 
through the photosphere and to fill the empty space outside it.
This radiation has a non-zero angular component of the net flux 
coming from the torus rotation (radiation is emitted 
isotropically in the comoving frame of optically thick gas). 
It initially reaches the polar axis, and then starts to interact 
with radiation emitted from the other side of the torus.
Effectively, the emitted radiation converge towards the axis.
The M1 scheme assumes the specific intensity is symmetric with
respect to the net flux. Therefore, it cannot mix colliding
beams of light properly. In the case of torus radiation interacting
near the polar axis, its horizontal component of flux vanishes
as a result of superposition of oppositely oriented beams. The sub-dominant 
vertical component sums up, as well as the azimuthal component.
The resulting specific intensity follows the net flux which is
dominated by the azimuthal component. It turns out that it is
so collimated along $\phi$ that the radiation does not reach
the polar axis. 

\begin{figure}
\includegraphics[width=1.\columnwidth]{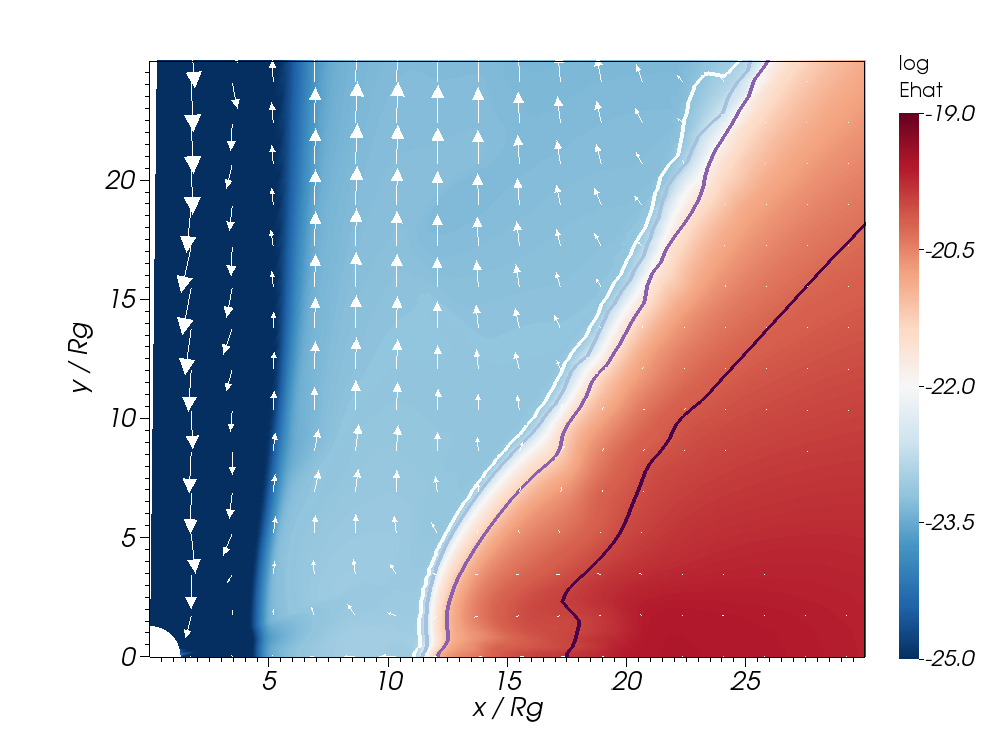}\\
 \includegraphics[width=1.\columnwidth]{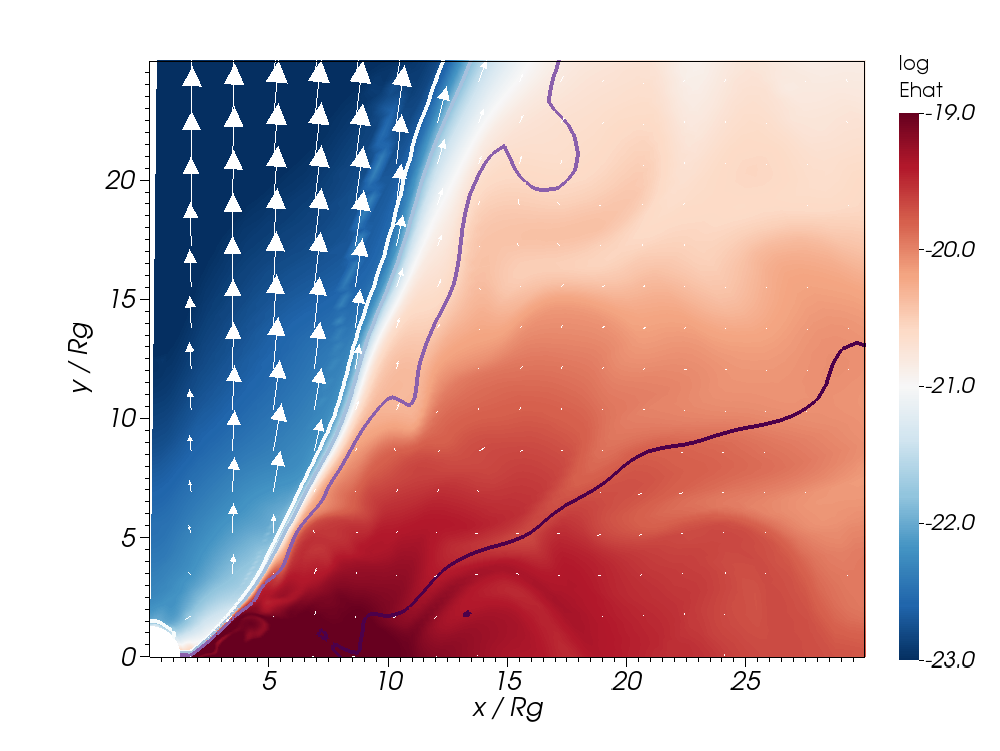}
\caption{Comoving frame radiative energy density ($\widehat E$) for 
  $a_*=0.9$ at $t=750$ (top) and $t=8000$ (bottom). Arrows reflect
the radiation rest frame velocity on the poloidal plane. Contours
show the total optical depth and range from $\tau=1$ (white) to
$\tau=1000$ (deep violet). The bottom panel corresponds to the left top-left half of
the middle panel in Fig.~\ref{f.400a9_triple}.}
  \label{f.400a9_shock}
\end{figure}

This effect is shown 
in the top panel of Fig.~\ref{f.400a9_shock} which presents the radiative energy
density in the comoving frame ($\widehat E$) at an early
($t=750$) stage
of the $a_*=0.9$ simulation. Contours show the total
optical depth (white contour corresponds to $\tau=1$), and
the arrows denote the magnitude and direction of the 
radiation rest frame velocity on the poloidal plane (large velocities correspond
to a highly collimated radiation in optically thin region). 
The red region is the optically
thick torus and reflects high energy density of trapped radiation.
The light blue region in the middle is the disk optically thin
atmosphere filled with radiation emerging from the torus surface and
propagating away from the midplane. The dark blue region
nearest to the polar axis is not accessible for the
torus radiation because of the effect described above. At the border,
a strong discontinuity in $\widehat E$ forms, and the polar region
is filled with residual radiation coming from near the outer edge.

This unphysical configuration takes place only in the initial stage.
Once MRI develops and torus is turbulent, it adjusts to the equlibrium
solution: it extends down to the BH, the photosphere moves closer to the polar axis,
and the rotation becomes sub-Keplerian. As the bottom panel shows,
such a configuration allows the radiation to reach the axis, and
the discontinuity disappeares. Therefore, the artificial radiation
shock is not affecting the results described in this paper (still, the
radiation distribution in the atmosphere is only qualitatively right
because of approximate nature of any closure scheme).
However, accretion disks with lower accretion rates, which rotate faster and
have photosphere further out from the polar axis, will likely produce
such artificial discontinuities. We will tackle this problem in a forthcoming paper.

\end{document}